\documentclass[conference]{IEEEtran}

\usepackage{graphicx}
\usepackage{epstopdf}
\usepackage{standalone}
\usepackage[nolist ]{acronym}
\usepackage{tcolorbox}

\usepackage{pdfpages}
\usepackage{tikz}
\usetikzlibrary{positioning}
\usepackage{hyperref}
\usepackage{pdfcomment}
\usepackage{hologo}

\usepackage{xstring}

\usepackage[utf8]{inputenc}
\usepackage{marvosym}

\usepackage{algorithm}
\usepackage{algpseudocode}
\usepackage{tikz}

\usepackage{listings}
\definecolor{ListingBackground}{rgb}{0.97,0.97,0.97}
\usepackage{pgfplots}
\pgfplotsset{compat=newest}
\pgfplotsset{
	box plot/.style={
		/pgfplots/.cd,
		fill=blue!30,
		only marks,
		mark=-,
		mark size=0.2em,
		/pgfplots/error bars/.cd,
		y dir=plus,
		y explicit,
	},
	box plot box/.style={
		/pgfplots/error bars/draw error bar/.code 2 args={%
			\draw  ##1 -- ++(.2em,0pt) |- ##2 -- ++(-.2em,0pt) |- ##1 -- cycle;
		},
		/pgfplots/table/.cd,
		y index=2,
		y error expr={\thisrowno{3}-\thisrowno{2}},
		/pgfplots/box plot
	},
	box plot top whisker/.style={
		/pgfplots/error bars/draw error bar/.code 2 args={%
			\pgfkeysgetvalue{/pgfplots/error bars/error mark}%
			{\pgfplotserrorbarsmark}%
			\pgfkeysgetvalue{/pgfplots/error bars/error mark options}%
			{\pgfplotserrorbarsmarkopts}%
			\path ##1 -- ##2;
		},
		/pgfplots/table/.cd,
		y index=4,
		y error expr={\thisrowno{2}-\thisrowno{4}},
		/pgfplots/box plot
	},
	box plot bottom whisker/.style={
		/pgfplots/error bars/draw error bar/.code 2 args={%
			\pgfkeysgetvalue{/pgfplots/error bars/error mark}%
			{\pgfplotserrorbarsmark}%
			\pgfkeysgetvalue{/pgfplots/error bars/error mark options}%
			{\pgfplotserrorbarsmarkopts}%
			\path ##1 -- ##2;
		},
		/pgfplots/table/.cd,
		y index=5,
		y error expr={\thisrowno{3}-\thisrowno{5}},
		/pgfplots/box plot
	},
	box plot median/.style={
		/pgfplots/box plot
	},
	boxplot/every median/.style={
		ultra thick,dashed,cyan
	}
}

\definecolor{flexicolor}{RGB}{46,49,146}
\definecolor{amaricolor}{RGB}{237,28,36}

\usepackage{xspace}

\usepackage[binary-units=true]{siunitx}
\usepackage{psfrag}
\usepackage{graphicx}
\usepackage{tabularx,booktabs}
\usepackage{multirow}
\usepackage{rotating}

\renewcommand{\baselinestretch}{1}

\usepackage{multirow}

\usepackage[cmex10]{amsmath}
\usepackage[caption=false,font=footnotesize]{subfig}
\usepackage{stfloats}
\renewcommand{\vec}{\mathbf}	
\renewcommand{\baselinestretch}{0.96}	

\begin{document}

\newcommand{\paperTitle}{Lightweight Simulation of Hybrid Aerial- and Ground-based Vehicular Communication Networks}
\newcommand{\paperAuthors}{Benjamin Sliwa, Manuel Patchou, and Christian Wietfeld}
\newcommand{\paperEmails}{$\{$Benjamin.Sliwa, Manuel.Mbankeu, Christian.Wietfeld$\}$@tu-dortmund.de}

\newcommand{\figurePadding}{0pt}
\newcommand{\figureTopPadding}{\figurePadding}
\newcommand{\figureBottomPadding}{\figurePadding}

\newcommand{\dummy}[3]
{
	\begin{figure}[b!]  
		\begin{tikzpicture}
		\node[draw,minimum height=6cm,minimum width=\columnwidth]{\LARGE #1};
		\end{tikzpicture}
		\caption{#2}
		\label{#3}
	\end{figure}
}

\newcommand{\wDummy}[3]
{
	\begin{figure*}[b!]  
		\begin{tikzpicture}
		\node[draw,minimum height=6cm,minimum width=\textwidth]{\LARGE #1};
		\end{tikzpicture}
		\caption{#2}
		\label{#3}
	\end{figure*}
}

\newcommand{\basicFig}[7]
{
	\begin{figure}[#1]  	
		\vspace{#6}
		\centering		  
		\includegraphics[width=#7\columnwidth]{#2}
		\caption{#3}
		\label{#4}
		\vspace{#5}	
	\end{figure}
}
\newcommand{\fig}[4]{\basicFig{#1}{#2}{#3}{#4}{0cm}{0cm}{1}}

\newcommand{\subfig}[3]
{
	\subfloat[#3]{\includegraphics[width=#2]{#1}}\hfill
}

\newcommand\subDummy[4]
{
	\subfloat[#4]{\begin{tikzpicture}\node[draw,minimum height=#2,minimum width=#3]{\LARGE #1};\end{tikzpicture}}\hfill
}

\newcommand\wD{0.23\textwidth}
\newcommand\wT{0.32\textwidth}
\begin{acronym}
	\acro{eNB}{evolved Node B}
	\acro{gNB}{Next Generation Node B}
	
	\acro{ITS}{Intelligent Transportation Sytem}
	\acro{LTE}{Long Term Evolution}
	\acro{MANET}{Mobile Ad-hoc Network}
	\acro{ns-3}{Network Simulator 3}
	\acro{OMNeT++}{Objective Modular Network Testbed in C++}
	\acro{UAV}{Unmanned Aerial Vehicle}
	\acro{VANET}{Vehicular Ad-hoc Network}
	\acro{FANET}{Flying Ad-hoc Network}
	\acro{IDM}{Intelligent Driver Model}
	\acro{MOBIL}{Minimizing Overall Braking Induced by Lane Changes}
	\acro{IPC}{Interprocess Communication}
	\acro{OSM}{OpenStreetMap}
	\acro{CBR}{Constant Bitrate}
	\acro{UDP}{User Datagram Protocol}
	\acro{KPI}{Key Performance Indicator}
	
	\acro{UI}{User Interface}
	\acro{HIL}{Hardware-in-the-loop}
	\acro{LIMoSim}{Lightweight ICT-centric Mobility Simulation}
	\acro{CUSCUS}{CommUnicationS-Control distribUted Simulator}
	\acro{FL-AIR}{Framework libre AIR}
	\acro{C-V2X}{Cellular Vehicle-to-Everything}
	\acro{OpenGL}{Open Graphics Library}
	\acro{DES}{Discrete Event Simulator}
	\acro{WGS84}{World Geodetic System 1984}
	
	\acro{UE}{User Equipment}
	\acro{eNB}{evolved Node B}
	\acro{RSRP}{Reference Signal Received Power}
	\acro{SINR}{Signal-to-interference-plus-noise Ratio}
	\acro{IPG}{Interpacket Gap}
	\acro{CAM}{Cooperative Awareness Message}
	\acro{PDR}{Packet Delivery Ratio}
	\acro{LOS}{Line-of-sight}
	\acro{NLOS}{Non-line-of-sight}
\end{acronym}

\newcommand\its{\ac{ITS}\xspace}
\newcommand\itss{\acp{ITS}\xspace}
\newcommand\uav{\ac{UAV}\xspace}
\newcommand\uavs{\acp{UAV}\xspace}
\newcommand\limosim{\ac{LIMoSim}\xspace}

\newcommand\fanet{\ac{FANET}\xspace}
\newcommand\fanets{\acp{FANET}\xspace}
\newcommand\ipc{\ac{IPC}\xspace}

\acresetall
\title{\paperTitle}

\author{\IEEEauthorblockN{\textbf{\paperAuthors}}
	\IEEEauthorblockA{Communication Networks Institute, TU Dortmund University,	44227 Dortmund, Germany\\ e-mail: \paperEmails}}

\maketitle

%
% Make your adjustments here
%
\def\COPYRIGHTYEAR{2019}
\def\CONFERENCE{2019 IEEE 90th IEEE Vehicular Technology Conference (VTC-Fall)} % set after the paper has been accepted
\def\DOI{10.1109/VTCFall.2019.8891340}	% set after the paper has been published

\def\bibtex
{
	@InProceedings\{Sliwa/Wietfeld/2019b,\\
	author    = \{Benjamin Sliwa and Christian Wietfeld\},\\
	title     = \{Empirical analysis of client-based network quality prediction in vehicular multi-\{MNO\} networks\},\\
	booktitle = \{2019 IEEE 90th Vehicular Technology Conference (VTC-Fall)\},\\
	year      = \{2019\},\\
	address   = \{Honolulu, Hawaii, USA\},\\
	month     = \{Sep\},\\
	\}
}
\ifx\CONFERENCE\VOID
\def\conferencenotice{Submitted for publication}
\def\copyrightnotice{}
\else
\ifx\DOI\VOID
\def\conferencenotice{Accepted for presentation in: \CONFERENCE}	
\else
\def\conferencenotice{Published in: \CONFERENCE\\DOI: \href{http://dx.doi.org/\DOI}{\DOI}

}
\fi
\def\copyrightnotice{
	\copyright~\COPYRIGHTYEAR~IEEE. Personal use of this material is permitted. Permission from IEEE must be obtained for all other uses, including reprinting/republishing this material for advertising or promotional purposes, collecting new collected works for resale or redistribution to servers or lists, or reuse of any copyrighted component of this work in other works.
}
\fi
\def\overlayimage{%
	\begin{tikzpicture}[remember picture, overlay]
		\node[below=5mm of current page.north, text width=20cm,font=\sffamily\footnotesize,align=center] {\conferencenotice \vspace{0.3cm} \pdfcomment[color=yellow,icon=Note]{\bibtex}};
	\node[above=5mm of current page.south, text width=15cm,font=\sffamily\footnotesize] {\copyrightnotice};
	\end{tikzpicture}%
}
\overlayimage
\begin{abstract}
		
%
% Introduction
%
Cooperating small-scale \uavs will open up new application fields within next-generation \itss, e.g., airborne near field delivery.
%
% Problem statement
%
In order to allow the exploitation of the potentials of hybrid vehicular scenarios, reliable and efficient bidirectional communication has to be guaranteed in highly dynamic environments.
%
% Solution appraoch
%
For addressing these novel challenges, we present a lightweight framework for integrated simulation of aerial and ground-based vehicular networks. Mobility and communication are natively brought together using a shared codebase coupling approach, which catalyzes  the development of novel context-aware optimization methods that exploit interdependencies between both domains.
%
% Results
%
In a proof-of-concept evaluation, we analyze the exploitation of \uavs as local aerial sensors as well as aerial base stations. In addition, we compare the performance of \ac{LTE} and \ac{C-V2X} for connecting the ground- and air-based vehicles.

\end{abstract}

\IEEEpeerreviewmaketitle
\section{Introduction}

%
% Introduction: UAVs in ITSs
%
The exploitation of the unique mobility characteristics of autonomous small-scale \uavs offers new possibilities for next generation \itss \cite{Sliwa/etal/2019b}, e.g., accident reporting, near field delivery, and network provisioning via \fanets and aerial base stations \cite{Menouar/etal/2017a}. An overview about multiple example use cases is illustrated in Fig.~\ref{fig:scenario}. As a consequence, \its-based mobility management and communication systems will be required to support heterogeneous vehicle classes, which raises new challenges for both domains \cite{Cheng/etal/2018a}.

%
% Problem statement
%
The development of novel systems and algorithms for addressing these challenges requires simulation frameworks, which are capable of modeling the mobility characteristics as well as the communication technologies for the different traffic participants \cite{Djahel/etal/2015a}. While the simulation of vehicular mobility has been widely studied \cite{Cavalcanti/etal/2018a} and different tools have been established, \uav networks have mostly been analyzed numerically or using highly-specialized simulators.

%
% Solution approach vs state of the art
%
In this paper, we present an open simulation framework for system-level modeling of \uav-enabled \itss, which brings together aerial and ground-based communication networks. The proposed simulator \limosim implements a shared codebase approach to provide a straightforward coupling with established network simulators (\ac{ns-3} and \ac{OMNeT++}). In contrast to state-of-the-art \ipc-based coupling methods, this approach explicitly supports the development of anticipatory \cite{Bui/etal/2017a} communication systems, which exploit mobility-related context information for the optimization of decision processes.

%
% Fig. Scenario
%
\fig{b}{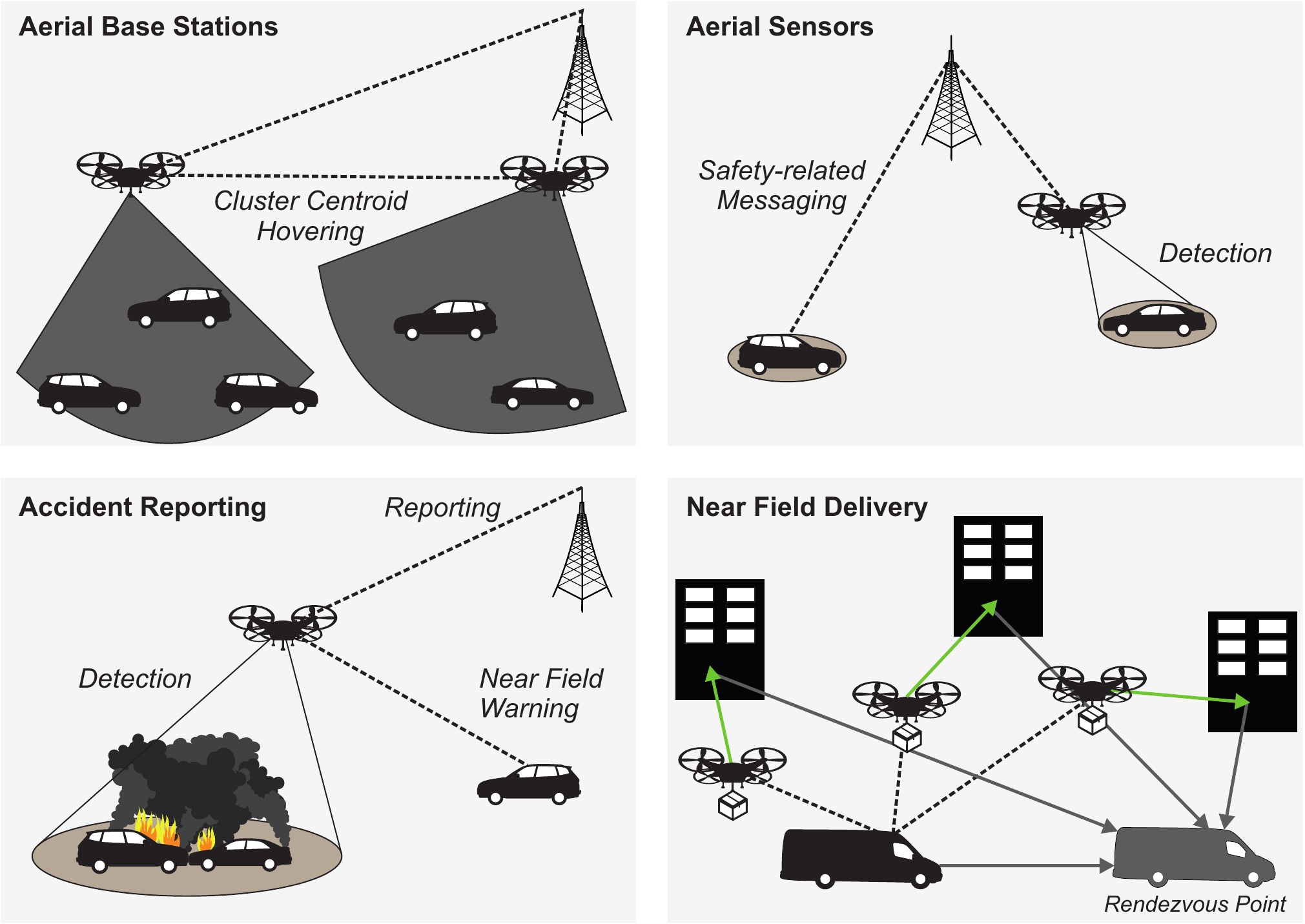}{Overview about example use cases of \uav-enabled \itss.}{fig:scenario}
%
% Contributions
%
Within this paper, the following contributions are provided:
\begin{itemize}
	\item Presentation of a novel framework for \textbf{joint simulation of aerial and ground-based vehicular communication networks}, which relies on established analytical models for simulating the low level mechanisms.
	\item Proof-of-concept evaluation in different \textbf{case studies}, e.g., comparison of \ac{LTE} and \ac{C-V2X}  for safety-critical messaging.
\end{itemize}
%
% Structure of the paper
%
The paper is structured as follows: After discussing relevant state-of-the-art approaches in Sec.~\ref{sec:related_work}, we present the system model of the proposed simulator as well as its key components in Sec.~\ref{sec:approach}. Afterwards, the methodological setup for the simulative performance evaluation and the results of different case-studies are presented and discussed in Sec.~\ref{sec:results}.

\section{Related Work} \label{sec:related_work}

%
% Fig. Architecture model
%
\begin{figure*}[t]  	
	\centering		  
	\includegraphics[width=0.8\textwidth]{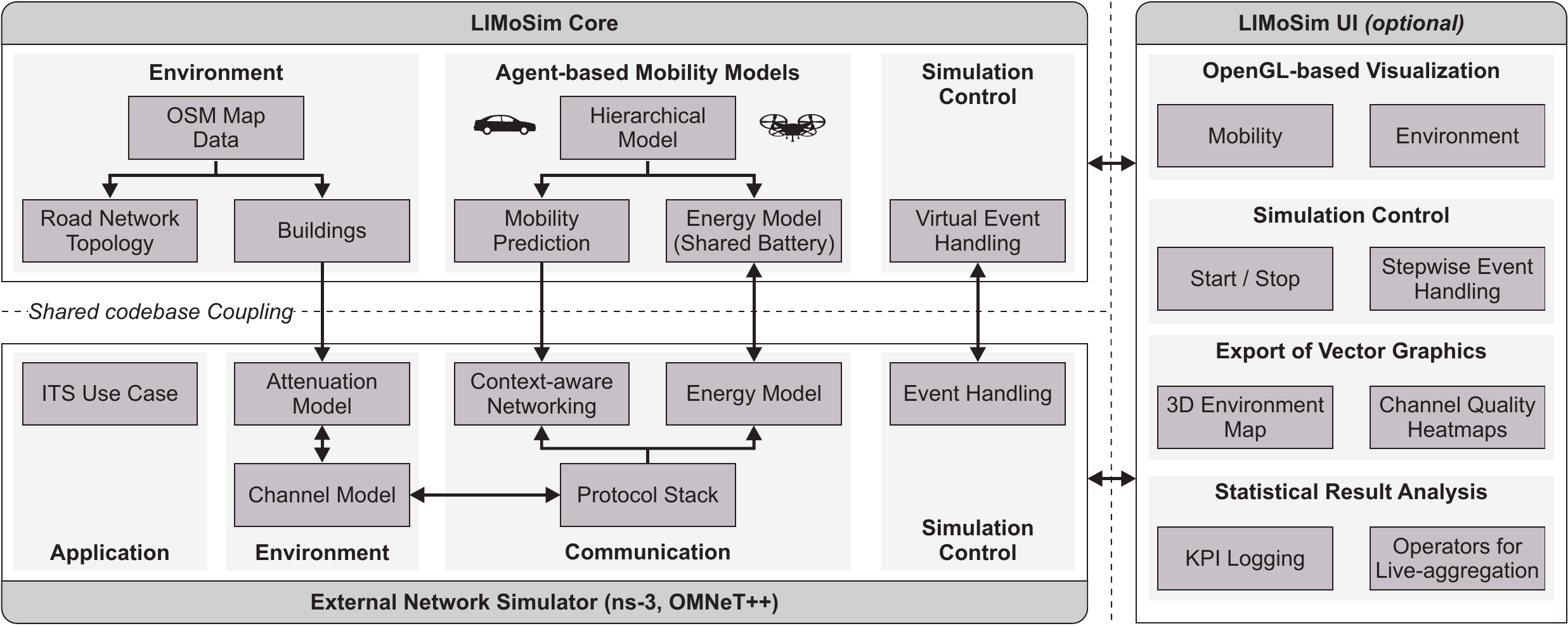}
	\caption{Overall system architecture model. The proposed \limosim consists of three main components and several abstract modules.}
	\label{fig:architecture}	
	\vspace{-0.6cm}
\end{figure*}
\uav networks are a subset of mobile robotic networks \cite{Cao/etal/2018a}, which rely on \emph{controlled mobility} to fulfill a specified task. An overview about commonly used mobility models is given in \cite{Xie/etal/2014a}.
%
% 3D Motion -> Impacts for Communication
%
While the capability for 3D motion offers the potential of overcoming limitations of ground-bound vehicles (e.g., for near field delivery \cite{Marinelli/etal/2018a}), the provision of reliable communication links is a challenging task due to the mobility-related dynamics of the radio channel. 
In order to proactively address these challenges, the \emph{anticipatory communication paradigm} \cite{Bui/etal/2017a} has been proposed, which exploits context information (e.g., mobility prediction) for optimizing communication processes.
%
% Anticipatory Networking 
%
In previous work, we have demonstrated that predictive methods can highly increase the quality of mesh routing decisions within \uav swarms by proactive consideration of their mobility characteristics \cite{Sliwa/etal/2016a, Sliwa/etal/2019a}. Furthermore, we have shown that the resource-efficiency of vehicular sensor data transmissions can be massively increased by applying cross-layer data rate prediction that exploits channel quality indicators as well as mobility information \cite{Sliwa/etal/2018a, Sliwa/etal/2018b, Sliwa/etal/2019d}.
%
% Communication-aware Mobility
%
A complementary approach for addressing the same challenges is the usage of \emph{communication-aware mobility} principles, which aim to proactively avoid challenging radio channel conditions by exploiting the mobility characteristics of the \uavs \cite{Goddemeier/etal/2012a} .
%
% Motiviation for Shared Codebase
%
Since all of these optimization approaches rely on a direct interaction of mobility and communication, the proposed \limosim models both components natively in a single process. The shared codebase approach allows to easily implement \emph{cross-layer optimization} methods that consider both domains for improving the overall system.

%
% Energy Limitations
%
The operation lifetime of small-scale \uavs is highly limited by the available energy resources. Although joint optimization of communication and trajectory design \cite{Zeng/Zhang/2017a} has become one of the major research fields of \uav networks, most analyzations are performed numerically, which limits the significance of the developed methods for complex scenarios.

%
% UAV Simulation Frameworks
%
While different simulation frameworks for \uav networks have been proposed, established approaches have highly specialized focus fields.
%
% FlyNetSim
%
FlyNetSim \cite{Baidya/etal/2018a} is a simulation tool that couples the simulators ArduPilot and \ac{ns-3} using a dedicated middleware. The simulator is designed agent-centric with a focus on \ac{HIL} evaluations. 
%
% CUSCUS
%
\ac{CUSCUS} \cite{Zema/etal/2018a} is a highly complex simulation framework, which interconnects \ac{ns-3} and the \ac{FL-AIR} mobility simulation using Linux containers. Due to the coupling approach, the provided support of communication technologies is limited, e.g., \ac{LTE} is currently not supported.
%
% Trace based approaches
%
Other methods omit explicit modeling of \uav mobility by relying on real world traces  \cite{Hadiwardoyo/etal/2019a}.
%
% LIMoSim vs Other Simulatior
%
In contrast to the state-of-the-art approaches, the proposed \limosim approach couples mobility and communication middleware-less, which results in a less complex setup phase for the creation of evaluation scenarios. Similar to well-known microscopic models of car mobility, the mobility behavior of the airborne vehicles is modeled from a system-level perspective.

\section{Overview About the \limosim Simulation Framework} \label{sec:approach}

The overall architecture model of the proposed simulation approach is illustrated in Fig.~\ref{fig:architecture}. 
%
% General Structure and Abstract Treatment of Network Simulators
%
\limosim consists of the main modules \emph{Core}, \emph{\ac{UI}} and a coupled external network simulator. The \emph{Core} provides the actual models for the mobility simulation and the environment representation. In order to enable shared codebase coupling with a wide range of different network simulators (e.g., \ac{ns-3} and \ac{OMNeT++}), the pure \texttt{C++}-based \emph{Core} does not have any dependencies towards the \texttt{Qt}-based \emph{\ac{UI}} or any network simulator. 

In the following paragraphs, the key modules of the simulation core \emph{mobility modeling}, \emph{energy modeling}, \emph{environment representation} and \emph{event synchronization} are further explained.
%
% Focus on UAV Networks
%
Due to spacial constraints, this paper mainly focuses on the aspect of modeling the \uav mobility. The models for the ground-based vehicles are further discussed in \cite{Sliwa/etal/2017a}.

\subsection{Coupling between Mobility and Communication}

\limosim and the respective network simulators are coupled in a shared codebase way. Therefore, modules from both domains are able to directly access and modify data in modules of the other domain.
%
% Synchronization
%
In order to guarantee causality between different events, the different \acp{DES} need to be synchronized. The shared codebase approach is hereby exploited to minimize the corresponding overhead. \limosim makes use of a \emph{virtual event queue}, which embeds its events to surrogates events in the domain of the coupled network simulator. The latter is then responsible for processing the event handling.
As soon as a surrogate event is called, it invokes the handler method of its mapped \limosim event.

\subsection{Online Visualization}

Visual verification is an important method during the design phase of novel mobility algorithms. Unfortunately,
the capability of network simulators to provide online visualization differs, e.g., \ac{ns-3} does not natively provide visualization features.
%
% OpenGL
%
Therefore, \limosim features a rich \ac{OpenGL}-based 3D visualization, which is implemented in \texttt{Qt-C++} and compatible with established operating systems. Note that this feature is optional and does not introduce any code dependencies to the core implementation, which is purely written in \texttt{C++}.

%
% EPS Export
%
In addition to live visualization, perspectival scenario figures can be exported in a vector data format (see Fig.~\ref{fig:map} as an example).

\subsection{Environment Modeling}

%
% OSM Support
%
In addition to generic environment models, the proposed simulation framework provides native support for \ac{OSM} data, which is utilized to automatically create the road network topology, buildings and infrastructure-based entities such as traffic signals. In contrast to other simulators, the data is parsed without requiring preprocessing steps with external tools.
%
% Optimization
%
The \texttt{*.osm} files are automatically converted to \texttt{*.limo} equivalents when they are loaded for the first time. The latter data format consists of only the required information and pre-computed values for the coordinates, which are transformed from \ac{WGS84} to their cartesian equivalents. This way, future evaluations benefit from a reduced scenario initialization time.
%
% Fig. Obstacle Modeling
%
\fig{}{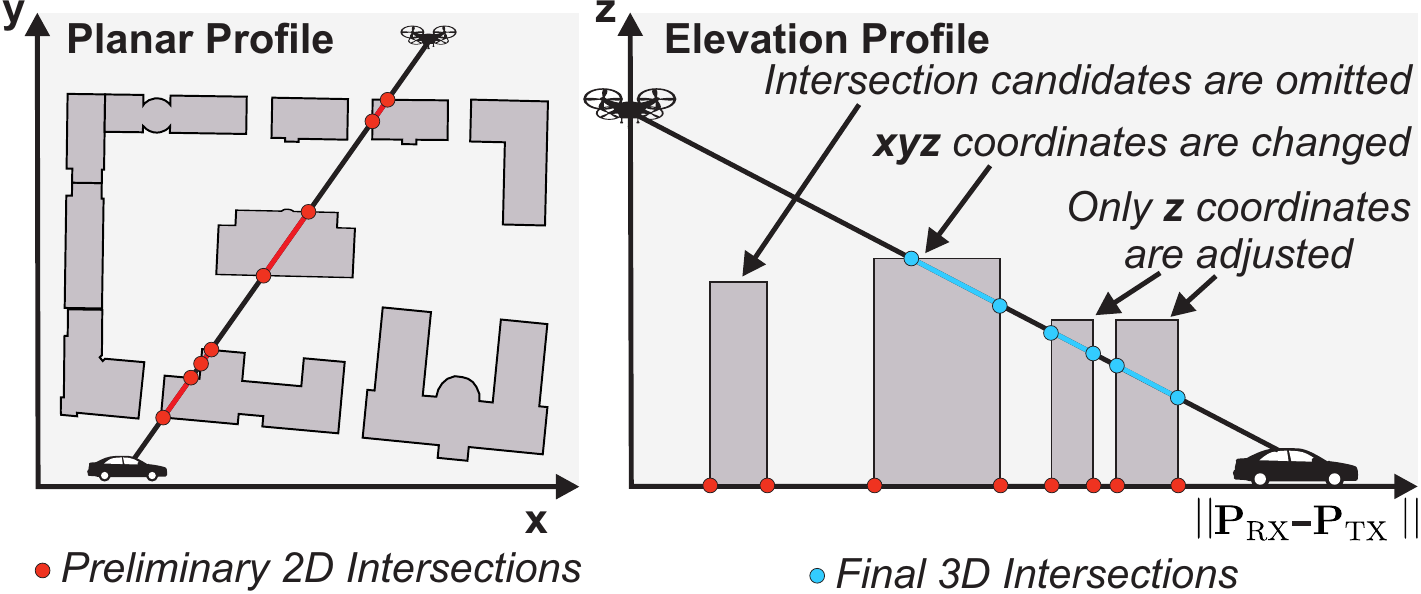}{Deterministic obstacle shadowing model. In the first step, intersection candidates are determined in the two-dimensional space. Afterwards, the coordinate system is transformed for considering the impact of the height information.}{fig:ray_tracing}

For considering shadowing-related attenuation in a deterministic way, \limosim provides a three-dimensional attenuation model, which exploits the building information contained in the \ac{OSM} data. The general approach is illustrated in Fig.~\ref{fig:ray_tracing}. At first, intersection candidates on the direct link are computed in the x-y-domain. Afterwards, the elevation profile of the link is analyzed and the precomputed candidates are either adjusted or omitted. As a result, the attenuated distance $d_{\text{obs}}$ and the number of intersections $N$ is derived and can be exploited by obstacle-aware channel models (e.g., \cite{Sommer/etal/2011a}).
For a given distance $d$, the resulting path loss $L$ is then computed by an obstacle-independent path loss model $L_{\text{PL}}$ and an additional obstacle attenuation term $L_{\text{obs}}$ as
\begin{equation}
	L = L_{\text{PL}}(d) + \underbrace{N\cdot \beta + d_{\text{obs}} \cdot \gamma}_{L_{\text{obs}}}
\end{equation}
with $\beta$ being the per wall attenuation and $\gamma$ being an intra-building attenuation factor. Fig.~\ref{fig:raytracing_time} shows an example of the resulting path loss dynamics for different receiver altitudes.

%
% Fig. Raytracing vs Time
%
\fig{}{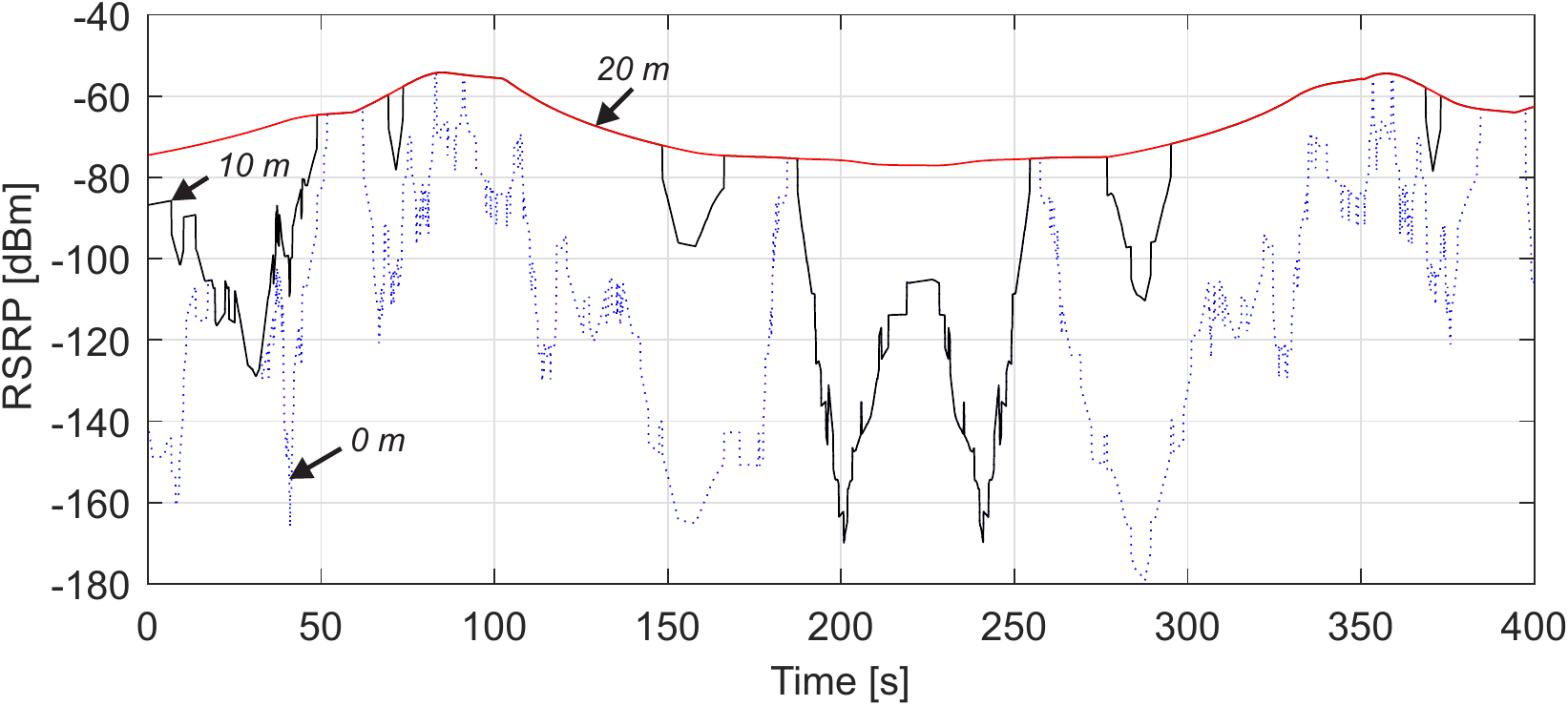}{Example temporal attenuation behavior of the obstacle shadowing model for different receiver altitudes.}{fig:raytracing_time}
%
%
%

%
% Car 
%
\subsection{Model of the Car Agent}

Further details of the car mobility implementation and its integration into \ac{OMNeT++} have been published in \cite{Sliwa/etal/2017a} and \cite{Sliwa/etal/2017b}. Similar to the \uav agent, the car agent relies on a hierarchical mobility model for considering different decision levels and control routines. As both vehicle class implementations are inherited from an abstract \texttt{Vehicle} base class, mobility-aware communication methods often do not require vehicle type-specific implementations, as both classes provide the same interfaces, e.g., for mobility prediction. 

As car following and lane change models, the classic \ac{IDM} and \ac{MOBIL} \cite{Treiber/etal/2000a} combination is implemented. On the \emph{strategic} layers, different random and deterministic route decision algorithms are available.

%
% UAV 
%
\subsection{Model of the \uav Agent}

As the aim of the proposed \limosim is to provide a development platform for novel methods, it makes use of a highly modular system architecture and abstract component definitions.

\subsubsection{Hierarchical Mobility Model}

%
% Fig. Reynolds
%
\basicFig{}{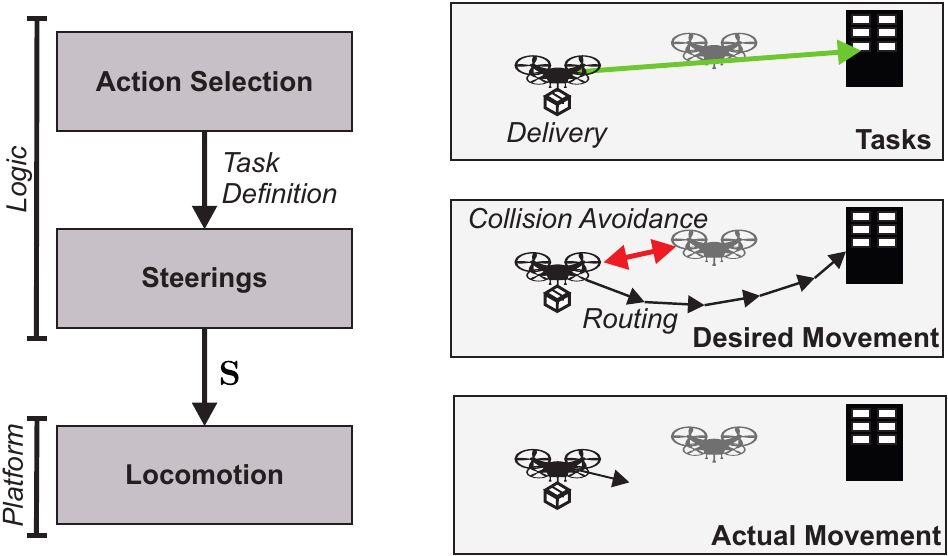}{Hierarchical mobility model inspired by \cite{Reynolds/1999a}, which is evaluated in each simulation step.}{fig:reynolds}{0cm}{0cm}{0.8}
%
% Reynolds Model
%
The general structure of the hierarchical mobility model is inspired by the theory about modeling swarm behaviors of \cite{Reynolds/1999a} and consists of the layers \emph{Action Selection}, \emph{Steerings} and \emph{Locomotion} (see Fig.~\ref{fig:reynolds}). The general idea of the model is to achieve a high grade of reusability of developed algorithms for different vehicle types. Therefore, \emph{logic} and \emph{implementation} are strictly separated. As a consequence, the developed algorithms may not only work with other (yet similar) vehicle types, they can also be easily transfered from the simulation domain to a real world platform, which then only needs to provide implementations for the lowest layers.

%
% Action Selection
%
The \emph{Action Selection} layer is used to determine the general behavior goals of the mobile agent. In order to guarantee seamless connectivity, communication-aware mobility behaviors often make use of different \emph{roles}, e.g., a subset of the \uavs explores a mission area while the remaining vehicles act as multi-hop relays to ensure a reliable connection to a remote base station \cite{Sliwa/etal/2016a}.

%
% Steerings
%
\emph{Steerings} are subroutines that execute a well-defined task, e.g., collision avoidance and waypoint following. They are closely related to the three fundamental swarm motion types \emph{separation}, \emph{cohesion} and \emph{alignment} \cite{Reynolds/1999a}.
In each update iteration, all steerings are executed sequentially. The output of each steering is a \emph{steering vector} $\vec{S}_{i}$, which represents the desired multidimensional acceleration of the steering. The overall steering vector $\vec{S}$, which is forwarded to the \emph{Locomotion} layer, is then computed as the weighted average of all individual steerings $\vec{S}_{i}$ and their corresponding weights $w_{i}$ as
%
% Eq. Steering Vector
%
\begin{equation} \label{eq:steering_vector}
	\vec{S} = \left( \sum_{i=0}^{N-1}  w_{i}\right)^{-1} \sum_{i=0}^{N-1} w_{i} \cdot \vec{S}_{i}
\end{equation}
%
% Locomotion
%
Within the \emph{Locomotion} stage, the desired movement vector is transfered into a \emph{traveled movement vector}, which is determined with respect to the vehicle's movement capabilities and its previous angular and linear accelerations.

%
% Locomotion Implementation
%
Per default, a simplified version of the proposed model of \cite{Luukkonen/2011a} is applied. With the angular vector $\eta = \begin{bmatrix} \theta \ \phi \ \psi \end{bmatrix}^{-1} $, which contains the values for \emph{pitch}, \emph{roll} and \emph{yaw}, the cartesian motion is computed as
\begin{equation}
	\begin{bmatrix} \ddot{x} \\ \ddot{y} \\ \ddot{z} \end{bmatrix} =
	-g \begin{bmatrix} 0 \\ 0 \\ 1 \end{bmatrix}
	+ \frac{T}{m}
	\begin{bmatrix} C_{\psi}S_{\theta}C_{\phi} + S_{\psi}S_{\phi} \\ S_{\psi}S_{\theta}C_{\phi} - C_{\psi}S_{\phi} \\ C_{\theta}C_{\phi} \end{bmatrix}
\end{equation}
where $C_{x}=\cos(x)$, $S_{x}=\sin(x)$, $g$ is the gravitation, $T$ is the thrust and $m$ the vehicle mass. In addition, a differential equation is used to determine the angular accelerations as 
\begin{equation}
	\mathbf{\ddot{\eta}} = \mathbf{J}^{-1}(\mathbf{\tau}_{B}-\mathbf{C}(\mathbf{\eta},\mathbf{\dot{\eta}})\mathbf{\dot{\eta}})
\end{equation}
where  $\mathbf{J}$ is a jacobian matrix, which maps the angular velocities to angular accelerations, $\mathbf{\tau}_{B}$ being the torque and $\mathbf{C}(\mathbf{\eta},\mathbf{\dot{\eta}})$ being the coriolis term, which contains gyroscopic and centripetal effects.

\subsubsection{Mobility Prediction}

%
% Mobility Prediction
%
As discussed in Sec.~\ref{sec:related_work}, mobility prediction is one of the key enabling methods for anticipatory communication. Therefore, \limosim natively implements different models for predicting future vehicle locations.
%
% Hierarchical prediction model
%
The hierarchical mobility prediction is an iterative process (see Fig.~\ref{fig:mobility_prediction}), which uses the most precise prediction method from all available methods within each iteration $i$. In the first iteration, the aggregated steering vector $\vec{S}$ is directly used, as it already provides a prediction for the desired location. Afterwards, waypoint information is utilized if available. As a fallback mechanism, a pure extrapolation-based approach is applied.

With the distance increment $d_{\text{inc}}=v\cdot\tau$, the estimated position $\tilde{\vec{P}}_{i+1}$ in the next iteration $i+1$ is computed as 
%
% Eq. Mobility Prediction
%
\begin{equation} \label{eq:mobility_prediction} 
	\tilde{\vec{P}}_{i+1}=
		\begin{cases}
		%
		% Steering Vector
		%
		\vec{P}_{i} + \mathbf{S}_{i} & i=0 \\	
		%
		% Waypoints
		%
		\vec{P}_{i} + \frac{\vec{W}_{i}-\vec{P}_{i}}{||\vec{W}_{i}-\vec{P}_{i}||} \cdot d_{\text{inc}} & \vec{W}_{i}~\text{available}  \\ 
		%
		% Extrapolation
		%
		\vec{P}_{i} + \frac{1}{h} \sum_{j=i}^{h} \frac{\vec{P}_{j}-\vec{P}_{j-1}}{||\vec{P}_{j}-\vec{P}_{j-1}||}
		 &  \text{else} 
		\end{cases}	
\end{equation}
where $W_{i}$ is the current target waypoint and $h$ represents the number of considered previous positions.

%
% Fig. Mobility prediction
%
\fig{}{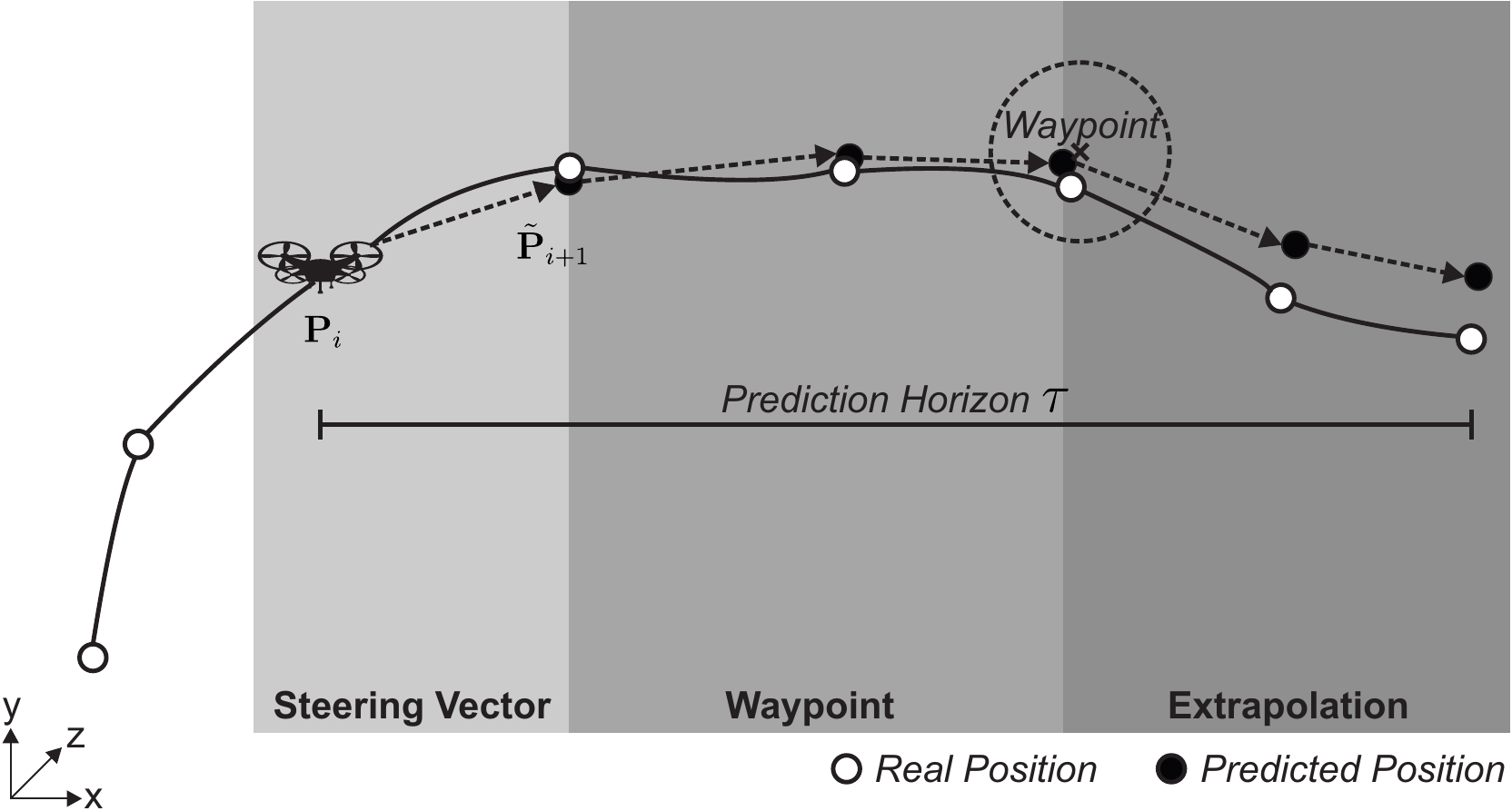}{Overview about the hierarchical mobility prediction model. The iterative processes utilizes the method with the highest available prediction accuracy in each iteration step.}{fig:mobility_prediction}

\subsubsection{Situation Awareness}

%
% Knowledge-base updated by Data Exchange -> can be exchanged by vision-based approaches
%
All vehicles maintain an awareness data base, which contains the positions of other nearby vehicles in order to enable collision avoidance. The information can be updated either distance-based (which mirrors vision-based approaches) or using message exchange.

\subsubsection{Power Consumption Model}

The overall operation time of the aerial vehicle is highly impacted by the mobility- and communication-related power consumption. The latter is depending on the communication pattern and the respective communication technology \cite{Liu/etal/2017a, Falkenberg/etal/2018a}. Within \limosim, the communication-related impact on the overall power consumption is computed based on available implementations provided by the supported network simulators.
%
% Logical Vehicle Battery
%
The mobility- and communication-related models are brought together in a logical vehicle battery.
%
% Coupling to the Power Consumption of the Communication
%
For the mobility side, the default implementation is inspired by the model proposed by \cite{Yacef/etal/2017a}.
%
% Fig. Energy vs Time
%
\begin{figure}[]  	
	\centering		  
	\includegraphics[width=1\columnwidth]{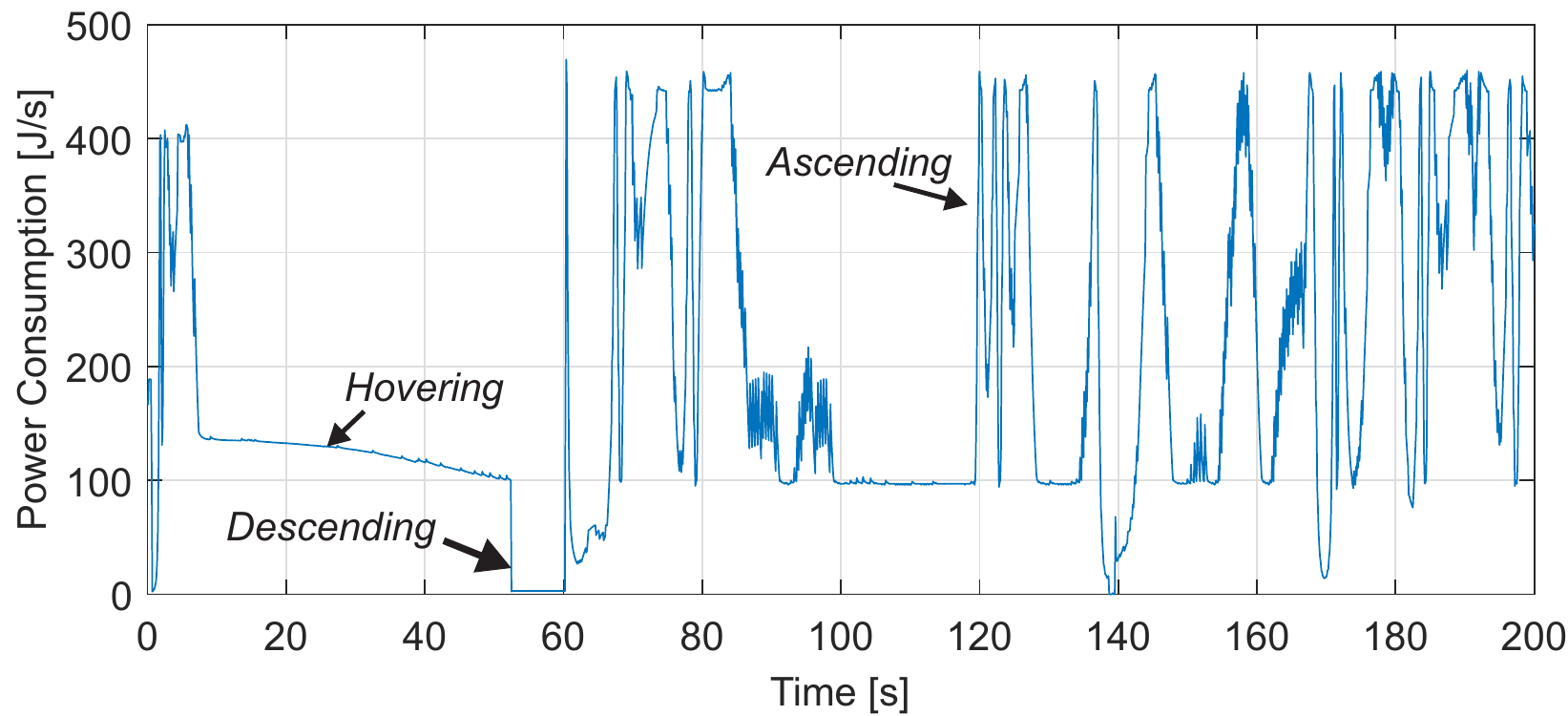}
	\vspace{-0.7cm}
	\caption{Example excerpt of the temporal behavior of the mobility-related power consumption behavior.}
	\label{fig:energy_time}
	\vspace{-0.5cm}	
\end{figure}
Fig.~\ref{fig:energy_time} shows an example excerpt of the temporal behavior of the mobility-related power consumption. Different characteristic motion states (\emph{ascend}, \emph{descend}, \emph{hover}) can be clearly identified. 

\section{Proof of Concept Evaluation} \label{sec:results} 

In this section, we present the setup and the results of different example case studies for hybrid vehicular communication networks.
%
% Fig. Map
%
\begin{figure}[]  	
	\centering		  
	\includegraphics[width=1\columnwidth]{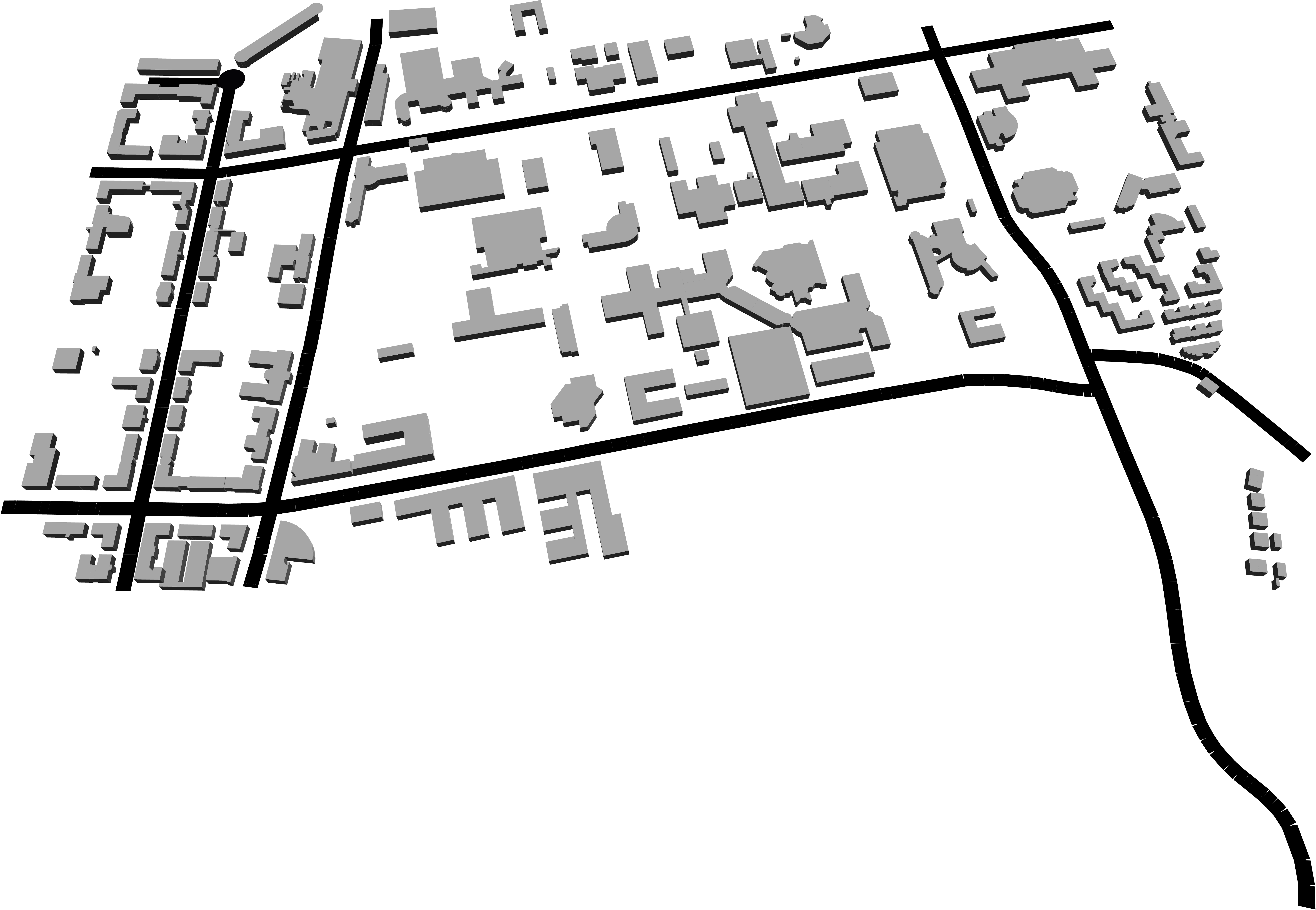}
	\vspace{-0.7cm}
	\caption{Aerial view on the reference scenario area. (Map data: ©OpenStreetMap contributors, CC BY-SA).}
	\label{fig:map}
	\vspace{-0.5cm}	
\end{figure}
The reference scenario is set up in a campus area, which is shown in Fig.~\ref{fig:map}. For the simulation of the communication behavior, \limosim is coupled with \ac{ns-3}.
%
% LTE vs C-V2X
%
As communication technologies, \ac{LTE} and \ac{C-V2X} (Mode~4) \cite{Eckermann/etal/2019a} are applied.
%
% Tab. Parameters
%
\newcommand{\entry}[2]{&#1 & #2 \\}
\newcommand{\head}[2]{& \toprule \entry{\textbf{#1}}{\textbf{#2}}}

\newcommand{\sideHeader}[3]
{
	\multirow{#1}{*}{
		\rotatebox[origin=c]{90}{
			\parbox{#2}{\centering \textbf{#3}}
		}
	}
}
										% TODO: Steerings + Locomotion
										
\begin{table}[ht]
	\centering
	\caption{Parameters for the ns-3 Simulation}
	\begin{tabular}{p{0.2cm}p{2.8cm}p{4.5cm}}
		\toprule
		\sideHeader{13}{0.5cm}{General} 
		\entry{\textbf{Parameter}}{\textbf{Value}}
		\midrule

		\entry{Simulation duration}{30~min}
		\entry{Runs per configuration}{10}
		\entry{Maximum speed (Car)}{14~mps}
		\entry{Maximum speed (\uav)}{20~mps}
		\entry{Scenario size}{1500~m x 750~m x 250~m}
		\entry{Channel model}{\textbf{HybridBuildingsPropagationLossModel}, DeterministicObstacleShadowing}

		\entry{Packet Size}{$\left\lbrace 100,~500,~\textbf{1000}..8000 \right\rbrace $~Byte}
		\entry{Inter Packet Interval}{$\left\lbrace \mathbf{10},~250,~500 \right\rbrace $~ms}

		%
		% V2X
		%
		\midrule
		\sideHeader{4.5}{0.3cm}{\mbox{C-V2X}} 
		\entry{Carrier frequency}{5.9~GHz}
		\entry{Bandwidth}{20~MHz}
		\entry{$P_{\text{TX}}$ (\acs{UE})}{23~dBm}
		
		%
		% LTE
		%
		\midrule
		\sideHeader{5}{0.3cm}{LTE} 
		\entry{Carrier frequency}{2.1~GHz}
		\entry{Bandwidth}{20~MHz}
		\entry{$P_{\text{TX}}$ (\acs{UE})}{23~dBm}
		\entry{$P_{\text{TX}}$ (\acs{eNB})}{43~dBm}

		\bottomrule
		
	\end{tabular}
	
	\label{tab:parameters}
	\vspace{0.1cm}
	$P_{\text{TX}}$: Maximum transmission power
\end{table}

A summary about the relevant parameters for the reference scenario is given in Tab.~\ref{tab:parameters}. All errorbars illustrate the 0.95-confidence interval of the mean value.

\subsection{\uavs as Aerial Sensors}

In the first case study, the \uavs are exploited as aerial sensors to increase the situation awareness of the cars. Each ground-based vehicle has an aerial \emph{sensor drone} (target height $40$~m), which periodically transmits sensor information as \acp{CAM} with an average packet size of $190$~Byte each $100$~ms. As these messages are potentially safety-related (e.g., collision warnings), the latency should be as low as possible. 
%
% Fig. Follower
%
\begin{figure}[] 
	\centering
	
	\subfig{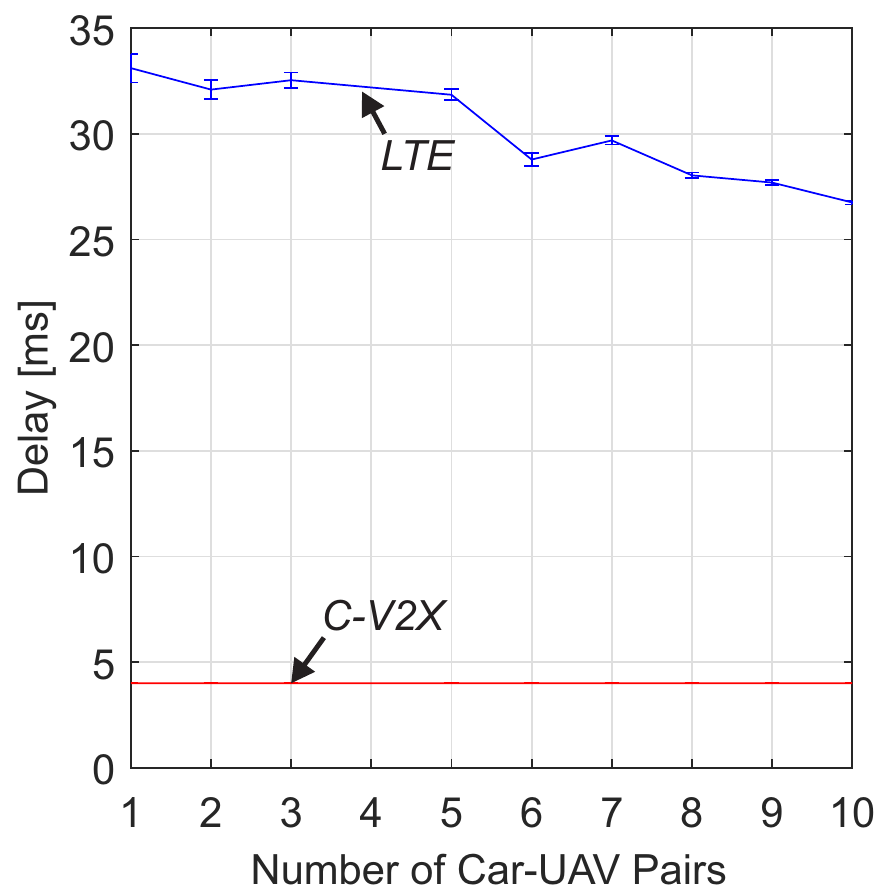}{\wD}{}
	\subfig{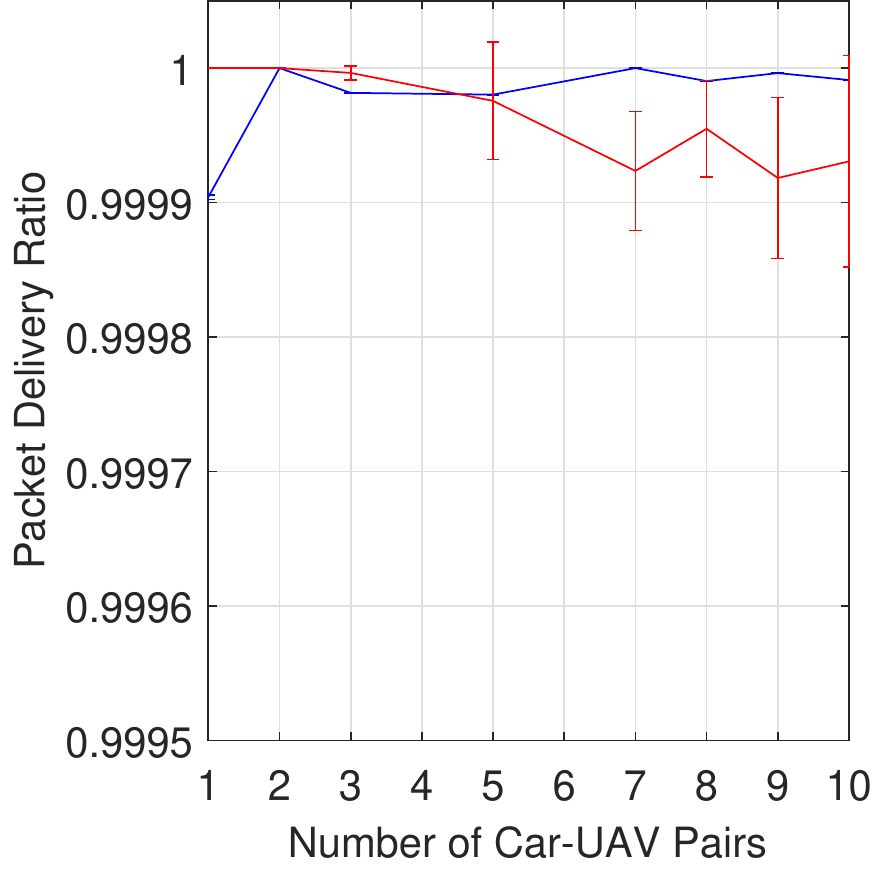}{\wD}{}
	
	\caption{Comparison between regular \ac{LTE} and \ac{C-V2X} for providing safety-related local awareness information.}
	\label{fig:follower}
	\vspace{-0.3cm}
\end{figure}

The statistical results for all evaluated simulation runs are shown in Fig.~\ref{fig:follower}. It can be seen that \ac{C-V2X} achieves a significantly smaller latency than \ac{LTE}. Since the applied variant of  \ac{C-V2X} implements the \emph{Mode~4} \cite{Seo/etal/2016a} behavior, the intra-\ac{UE} processing time has the dominant impact on the overall delay as the channel is accessed directly. In contrast to that, \ac{LTE} transmissions are impacted by the resource scheduling mechanism of the \ac{eNB}.

However, the direct channel access of \ac{C-V2X} increases the collision probability if larger numbers of \acp{UE} operate in the same interference region. As it can be seen in Fig.~\ref{fig:follower}~(b), the resulting \ac{PDR} is reduced with increasing \ac{UE} numbers.

\subsection{Providing Network Coverage with Aerial Base Stations}

%
% Motivation
%
In the second case study, the concept of aerial base stations (e.g., for providing high resolution map data for automated driving) is analyzed. In contrast to typical static deployments of the network infrastructure, the mobility capabilities of the \uavs offer the potential to dynamically react on shifts in the coverage requirements of the cell users. Within the \itss context, this approach allows to better react to spontaneous traffic phenomenons, which have an impact on the load of the cellular network, e.g., jam situations.
%
% Clusters
%
Within the simulation, a mobile aerial base station supplies a dynamic cluster of moving ground vehicles. Each vehicle requests a data stream as \ac{UDP} \ac{CBR} ($8000$~Byte each $10$~ms, which results in a requested traffic of $6.4$~MBit/s) in the downlink from a remote server.

%
% Fig. Aerial Base Station
%
\begin{figure}[] 
	\centering
	
	\subfig{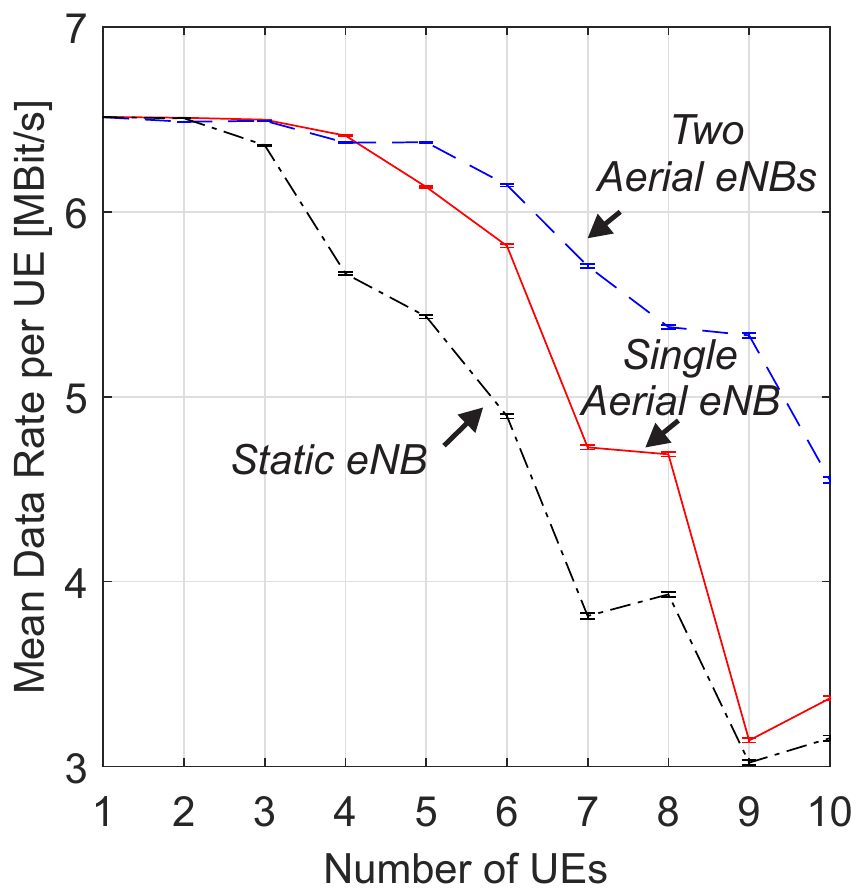}{\wD}{}
	\subfig{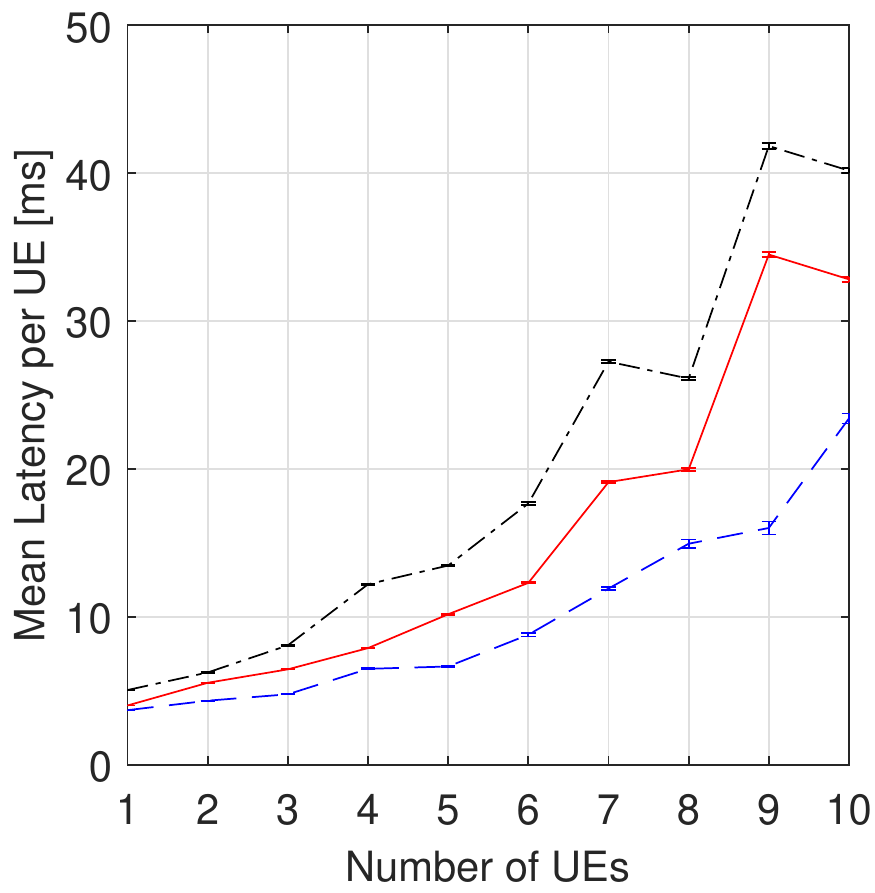}{\wD}{}
	
	\caption{Comparison of different \uav-enabled aerial base station variants and static \ac{eNB} deployments in the downlink direction.}
	\label{fig:aerial_bs}
	\vspace{-0.5cm}
\end{figure}
The results for mean data rate per \ac{UE} and mean delay per \ac{UE} for the air-to-ground link are shown in Fig.~\ref{fig:aerial_bs}. While the aerial base station approach is able to achieve significantly higher data rates than the static \ac{eNB} deployment, the achievable benefits are highly depending to the number of served \ac{UE}. 
For low amounts of \acp{UE}, the mobility of the \uav can be successfully exploited to increase the average network quality for all cell users.
For larger \ac{UE} amounts, the trajectories of the ground-based vehicles are too diverse for a single \uav to adopt to. As a consequence, the \uav behaves similar to a static \ac{eNB} in the center of the map. 
As shown in Fig.~\ref{fig:aerial_bs}, the degradation can be compensated by introducing additional \uavs with dedicated operation regions.

\subsection{Network Quality Prediction based on Connectivity Maps}

In the third case study, the suitability of the hierarchical mobility prediction scheme for forecasting the network quality at future locations is analyzed. Radio channel quality predictions can be exploited to enable context-aware communication strategies (e.g., opportunistic data transfer \cite{Sliwa/etal/2018a}).
%
% Fig. RSRP map
%
\basicFig{!h}{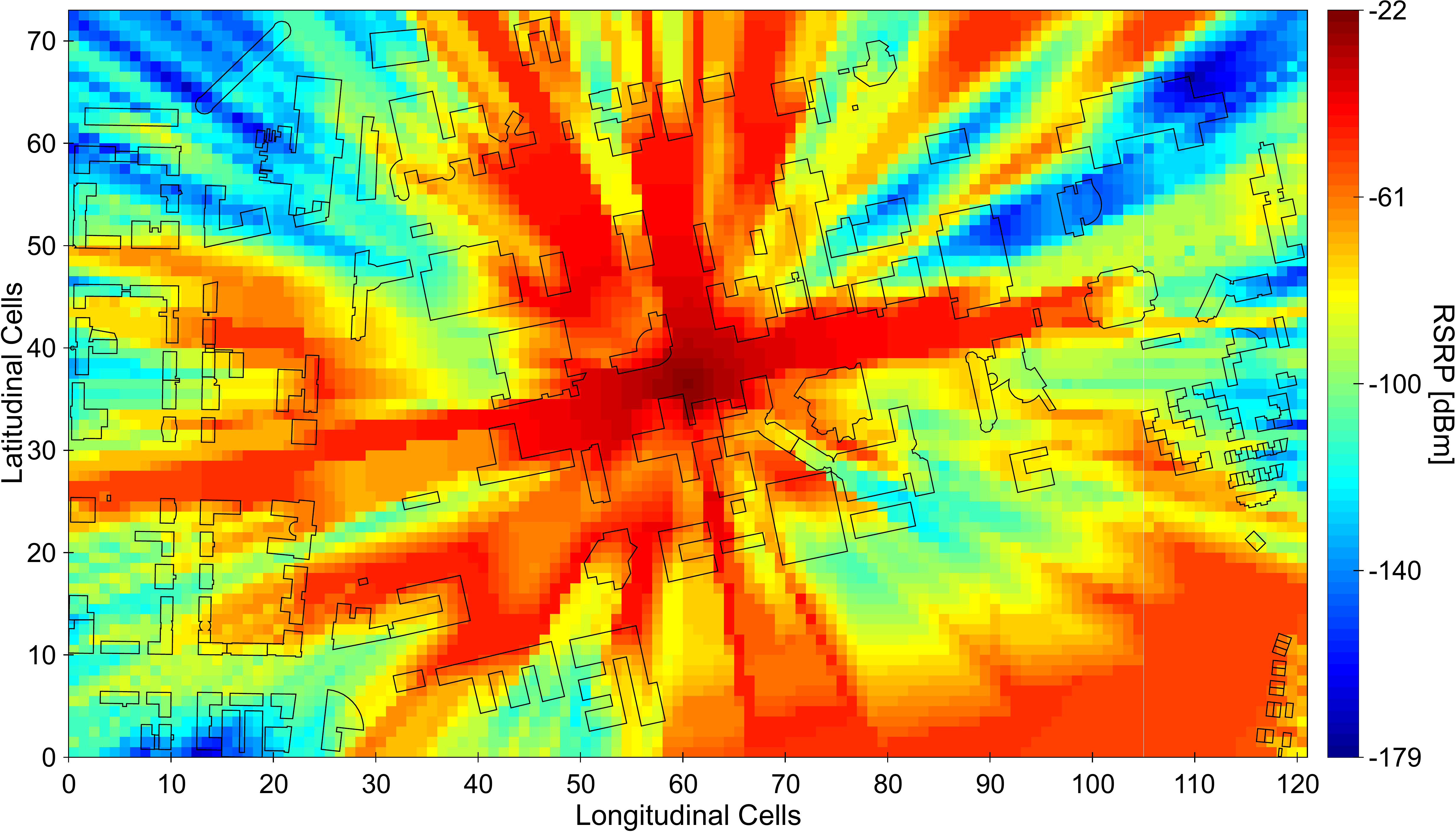}{Example \ac{RSRP} map (cell size 10~m) at ground height derived by the obstacle shadowing model.}{fig:rayTracing_map}{0cm}{0cm}{1}
In a preparatory step, a connectivity map for the \ac{RSRP} (see Fig.~\ref{fig:rayTracing_map}) is precomputed based on the determistic obstacle shadowing model and exploited as a priori information during the simulative evaluations. The \ac{eNB} height is 25~m and the buildings have random heights in the range of 10-30~m. Within each simulation run, the \uavs operate at a constant flight height and move based on random waypoints.
%
% Fig. Predictions
%
\begin{figure}[b] 
	\centering
	\vspace{-0.7cm}
	\subfig{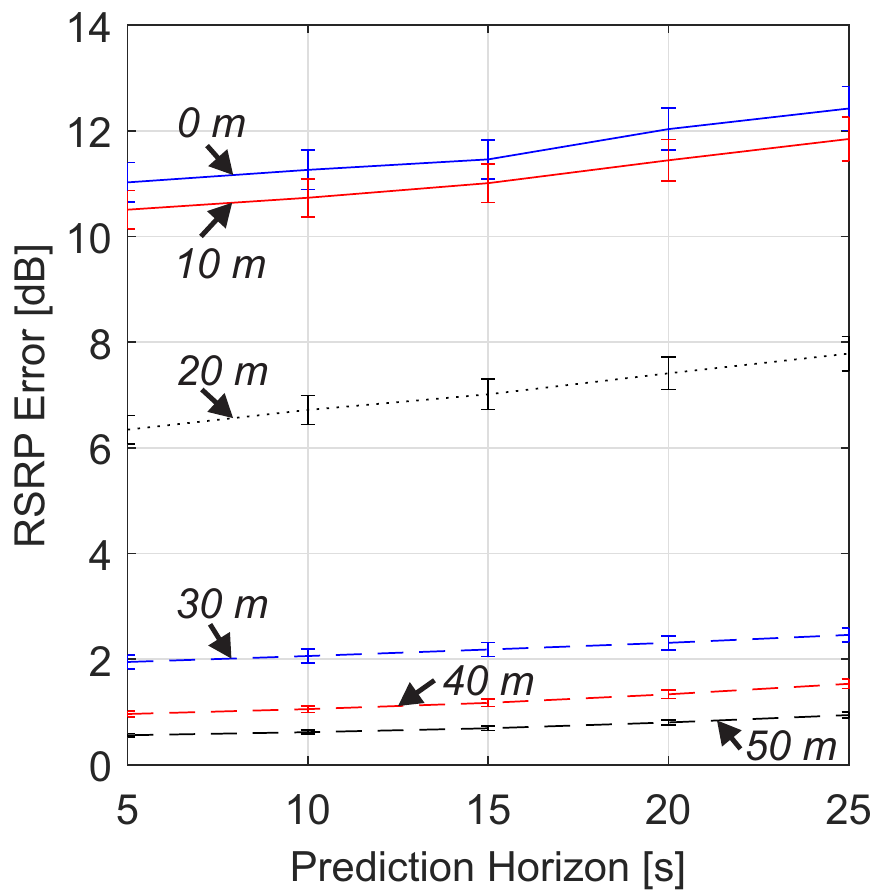}{\wD}{}
	\subfig{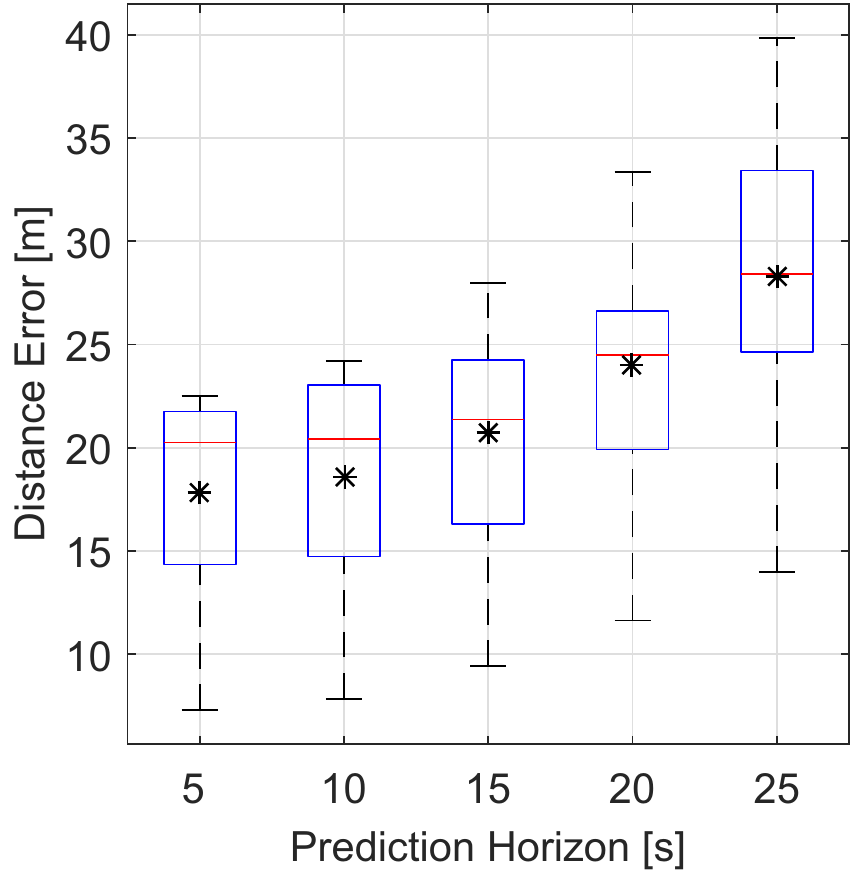}{\wD}{}
	
	\caption{Resulting prediction errors for different prediction horizons and flight altitudes.}
	\label{fig:prediction_errors}
\end{figure}
Fig.~\ref{fig:prediction_errors}~(a) shows the resulting prediction errors for the \ac{RSRP} for multiple flight altitudes. Three different areas can be identified. Up to 10~m, the building-related shadowing has a dominant impact on the resulting prediction error. While the absolute error of the position prediction (see Fig.~\ref{fig:prediction_errors}~(b)) is low, the remaining error range frequently leads to false \ac{LOS}~/~\ac{NLOS} decisions near the building corners.
At 20~m, the flight altitude is equal to the average building height. Therefore, \ac{LOS} and \ac{NLOS} situations occur equally often.
Above 30~m, the resulting \ac{RSRP} prediction error is very low, as only a few buildings are able to cause shadowing effects and the \uavs encounter \ac{LOS} situation most of the time.

\subsection{Scalability of the Simulator}

Finally, in order to assess the suitability of the proposed framework for large-scale simulation scenarios, the impact of the mobility simulation on the resulting computation time is analyzed.
All corresponding simulations are performed on an Intel Xeon X5690@3.47~GHz (4 cores), with 8GB RAM and Ubuntu 16.04 operating system with disabled \ac{UI}. 
%
% Fig. Scalability
%
\begin{figure*}[] 
	\centering
	
	\subfig{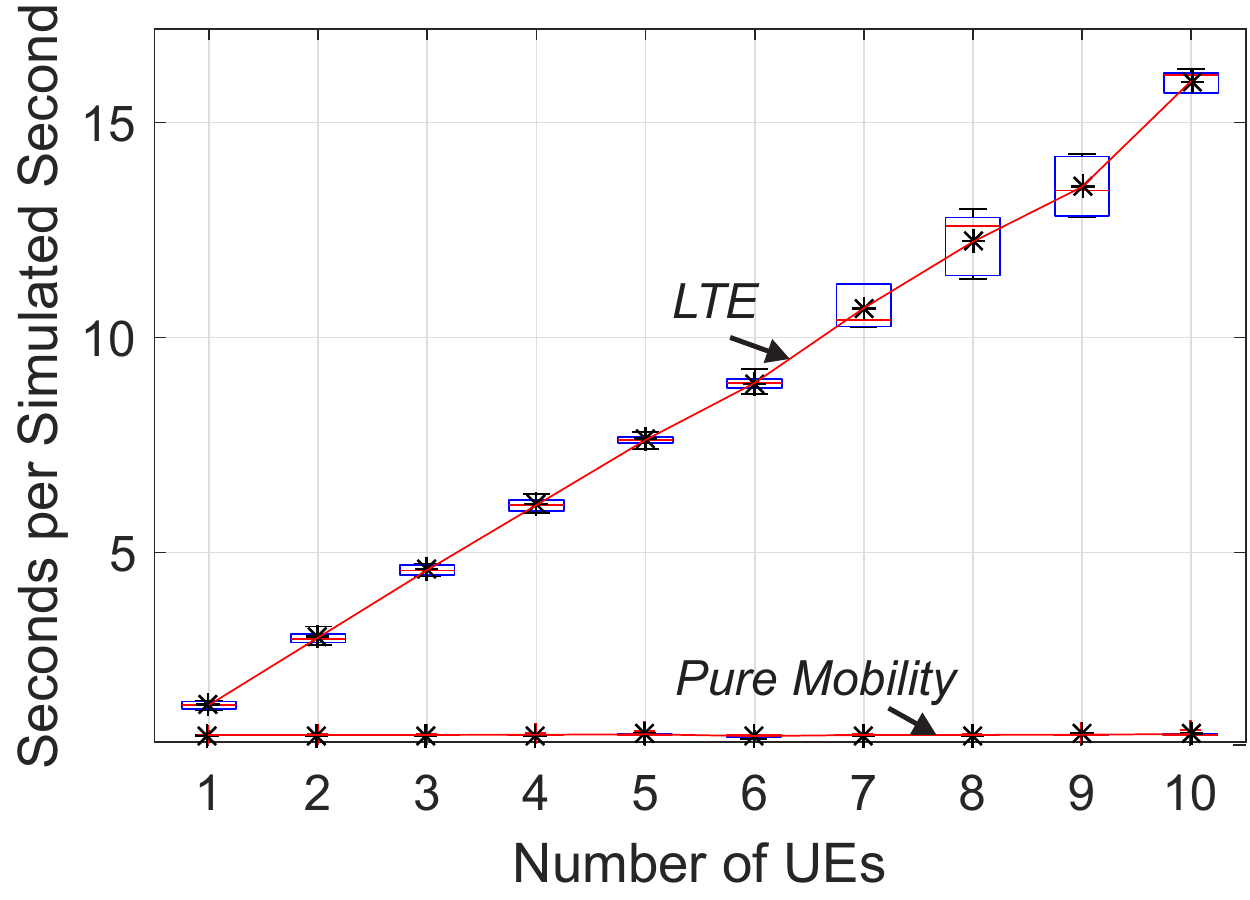}{\wT}{}
	\subfig{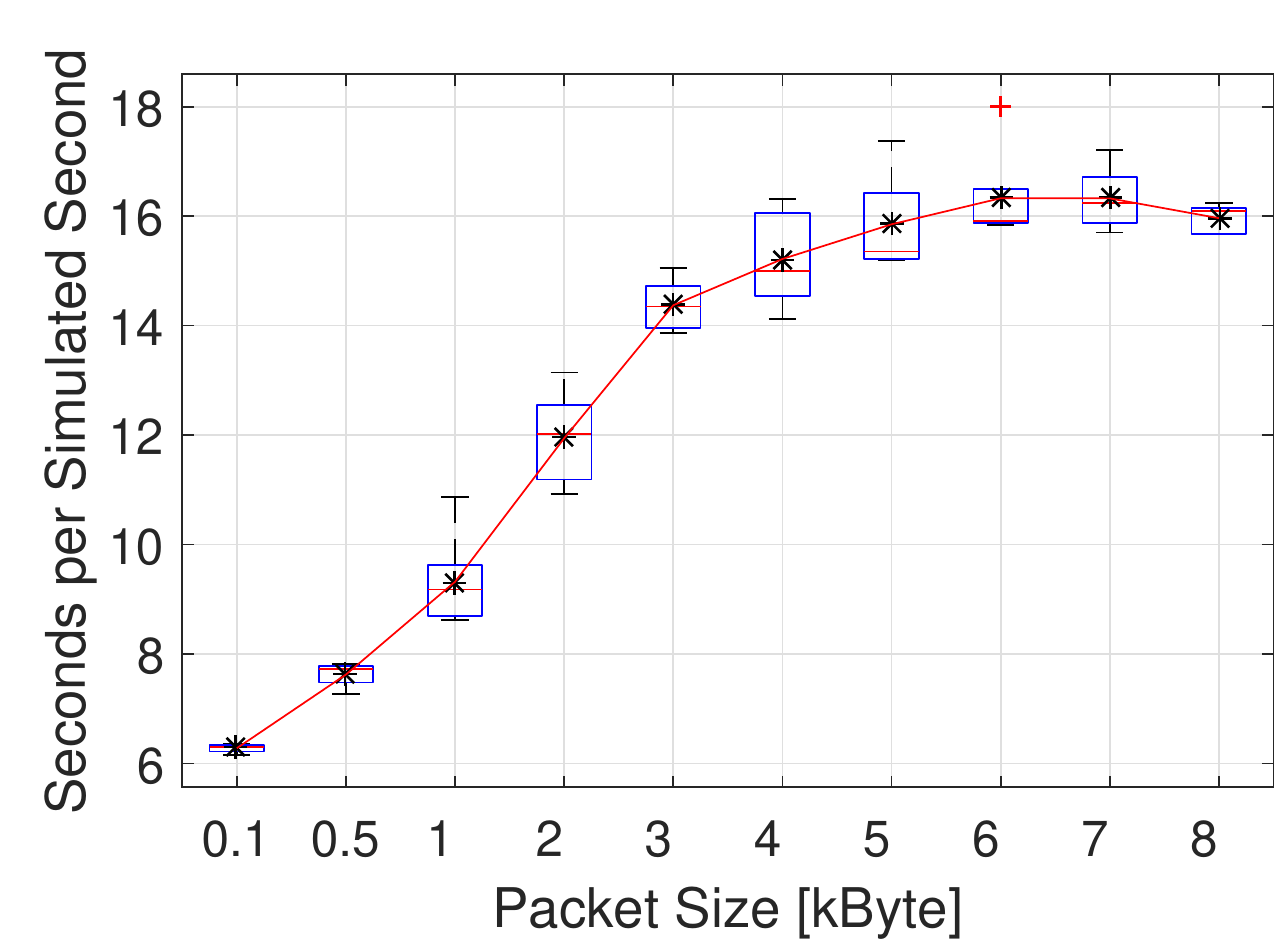}{\wT}{}
	\subfig{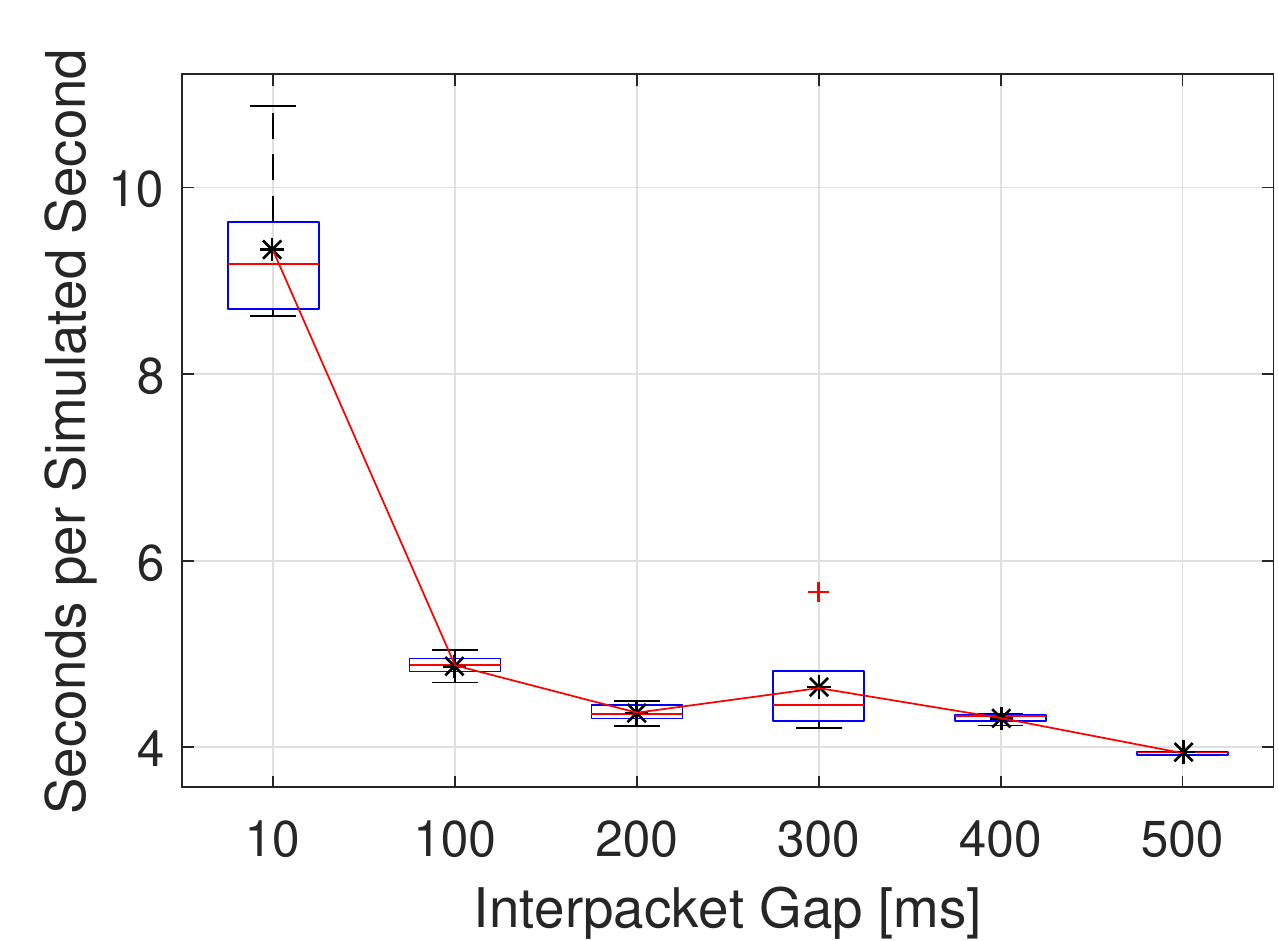}{\wT}{}

	\caption{Impact of different system dimensioning parameters on the resulting computation time (Aerial base station scenario).}
	\label{fig:scalability}
	\vspace{-0.7cm}
\end{figure*}
Fig.~\ref{fig:scalability} shows the normed resulting computation time with respect to the number of \acf{UE}, \ac{IPG} and packet size. 
It can be seen that the dominant runtime-related effects are caused by the network simulation, the impact of the mobility simulation by \limosim is almost negligible. The resulting behavior is linearly dependent to the number of modeled \acp{UE}. 

\section{Conclusion}

%
% Introduction
%
In this paper, we presented the open simulation framework \limosim, which provides a system-level development platform for joint modeling of aerial and ground-based vehicular networks.
%
% Problem statement
%
The proposed framework relies on an shared codebase coupling approach between mobility and communication simulation in order to catalyze the development of context-aware algorithms for future \uav-enabled \itss.
%
% Solution appraoch
%
While the physical motion and the energy consumption are represented by well-known analytical models, the hierarchical mobility model allows to seamlessly integrate novel steering routines.
%
% Results
%
As a proof-of-concept evaluation, the proposed simulator was applied to analyze the performance of safety-related messaging based on \ac{LTE} and \ac{C-V2X}. In addition, potential benefits of using aerial base stations for providing ground connectivity were evaluated.
%
% Future work
%
In future work, we will utilize \limosim to develop and evaluate novel anticipatory communication methods for aerial and ground-based vehicles. 
\section*{Acknowledgment}

\footnotesize
Part of the work on this paper has been supported by Deutsche Forschungsgemeinschaft (DFG) within the Collaborative Research Center SFB 876 ``Providing Information by Resource-Constrained Analysis'', project B4.

\bibliographystyle{IEEEtran}
\bibliography{Bibliography}

\begin{thebibliography}{10}
\providecommand{\url}[1]{#1}
\csname url@samestyle\endcsname
\providecommand{\newblock}{\relax}
\providecommand{\bibinfo}[2]{#2}
\providecommand{\BIBentrySTDinterwordspacing}{\spaceskip=0pt\relax}
\providecommand{\BIBentryALTinterwordstretchfactor}{4}
\providecommand{\BIBentryALTinterwordspacing}{\spaceskip=\fontdimen2\font plus
\BIBentryALTinterwordstretchfactor\fontdimen3\font minus
  \fontdimen4\font\relax}
\providecommand{\BIBforeignlanguage}[2]{{%
\expandafter\ifx\csname l@#1\endcsname\relax
\typeout{** WARNING: IEEEtran.bst: No hyphenation pattern has been}%
\typeout{** loaded for the language `#1'. Using the pattern for}%
\typeout{** the default language instead.}%
\else
\language=\csname l@#1\endcsname
\fi
#2}}
\providecommand{\BIBdecl}{\relax}
\BIBdecl

\bibitem{Sliwa/etal/2019b}
B.~Sliwa, T.~Liebig, T.~Vranken, M.~Schreckenberg, and C.~Wietfeld,
  ``System-of-systems modeling, analysis and optimization of hybrid vehicular
  traffic,'' in \emph{2019 Annual IEEE International Systems Conference
  (SysCon)}, Orlando, Florida, USA, Apr 2019.

\bibitem{Menouar/etal/2017a}
H.~Menouar, I.~Guvenc, K.~Akkaya, A.~S. Uluagac, A.~Kadri, and A.~Tuncer,
  ``{UAV}-enabled intelligent transportation systems for the smart city:
  {A}pplications and challenges,'' \emph{IEEE Communications Magazine},
  vol.~55, no.~3, pp. 22--28, March 2017.

\bibitem{Cheng/etal/2018a}
N.~Cheng, W.~Xu, W.~Shi, Y.~Zhou, N.~Lu, H.~Zhou, and X.~Shen, ``Air-ground
  integrated mobile edge networks: {A}rchitecture, challenges, and
  opportunities,'' \emph{IEEE Communications Magazine}, vol.~56, no.~8, pp.
  26--32, August 2018.

\bibitem{Djahel/etal/2015a}
S.~Djahel, R.~Doolan, G.~M. Muntean, and J.~Murphy, ``A communications-oriented
  perspective on traffic management systems for smart cities: c{}hallenges and
  innovative approaches,'' \emph{IEEE Communications Surveys Tutorials},
  vol.~17, no.~1, pp. 125--151, Firstquarter 2015.

\bibitem{Cavalcanti/etal/2018a}
E.~R. Cavalcanti, J.~A.~R. de~Souza, M.~A. Spohn, R.~C. d.~M. Gomes, and A.~F.
  B. F.~d. Costa, ``{VANETs}' research over the past decade: {O}verview,
  credibility, and trends,'' \emph{SIGCOMM Comput. Commun. Rev.}, vol.~48,
  no.~2, pp. 31--39, May 2018.

\bibitem{Bui/etal/2017a}
N.~Bui, M.~Cesana, S.~A. Hosseini, Q.~Liao, I.~Malanchini, and J.~Widmer, ``A
  survey of anticipatory mobile networking: Context-based classification,
  prediction methodologies, and optimization techniques,'' \emph{IEEE
  Communications Surveys \& Tutorials}, 2017.

\bibitem{Cao/etal/2018a}
X.~Cao, P.~Yang, M.~Alzenad, X.~Xi, D.~Wu, and H.~Yanikomeroglu, ``Airborne
  communication networks: {A} survey,'' \emph{IEEE Journal on Selected Areas in
  Communications}, vol.~36, no.~9, pp. 1907--1926, Sept 2018.

\bibitem{Xie/etal/2014a}
J.~Xie, Y.~Wan, J.~H. Kim, S.~Fu, and K.~Namuduri, ``A survey and analysis of
  mobility models for airborne networks,'' \emph{IEEE Communications Surveys
  Tutorials}, vol.~16, no.~3, pp. 1221--1238, Third 2014.

\bibitem{Marinelli/etal/2018a}
M.~Marinelli, L.~Caggiani, M.~Ottomanelli, and M.~Dell'Orco, ``En route
  truck-drone parcel delivery for optimal vehicle routing strategies,''
  \emph{IET Intelligent Transport Systems}, vol.~12, no.~4, pp. 253--261, 2018.

\bibitem{Sliwa/etal/2016a}
B.~Sliwa, D.~Behnke, C.~Ide, and C.~Wietfeld, ``{B.A.T.Mobile}: Leveraging
  mobility control knowledge for efficient routing in mobile robotic
  networks,'' in \emph{IEEE GLOBECOM 2016 Workshop on Wireless Networking,
  Control and Positioning of Unmanned Autonomous Vehicles (Wi-UAV)}, Washington
  D.C., USA, Dec 2016.

\bibitem{Sliwa/etal/2019a}
B.~Sliwa, S.~Falten, and C.~Wietfeld, ``Performance evaluation and optimization
  of {B.A.T.M.A.N. V} routing for aerial and ground-based mobile ad-hoc
  networks,'' in \emph{2019 IEEE 89th Vehicular Technology Conference
  (VTC-Spring)}, Kuala Lumpur, Malaysia, April 2019.

\bibitem{Sliwa/etal/2018a}
B.~Sliwa, T.~Liebig, R.~Falkenberg, J.~Pillmann, and C.~Wietfeld, ``Machine
  learning based context-predictive car-to-cloud communication using
  multi-layer connectivity maps for upcoming {5G} networks,'' in \emph{2018
  IEEE 88th Vehicular Technology Conference (VTC-Fall)}, Chicago, USA, Aug
  2018.

\bibitem{Sliwa/etal/2018b}
------, ``Efficient machine-type communication using multi-metric
  context-awareness for cars used as mobile sensors in upcoming {5G}
  networks,'' in \emph{2018 IEEE 87th Vehicular Technology Conference
  (VTC-Spring)}, Porto, Portugal, Jun 2018, {Best Student Paper Award}.

\bibitem{Sliwa/etal/2019d}
B.~Sliwa, R.~Falkenberg, T.~Liebig, N.~Piatkowski, and C.~Wietfeld, ``Boosting
  vehicle-to-cloud communication by machine learning-enabled context
  prediction,'' \emph{IEEE Transactions on Intelligent Transportation Systems},
  2019, {A}ccepted for publication.

\bibitem{Goddemeier/etal/2012a}
N.~Goddemeier, K.~Daniel, and C.~Wietfeld, ``Role-based connectivity management
  with realistic air-to-ground channels for cooperative {UAVs},'' \emph{IEEE
  Journal on Selected Areas in Communications}, vol.~30, no.~5, pp. 951--963,
  June 2012.

\bibitem{Zeng/Zhang/2017a}
Y.~Zeng and R.~Zhang, ``Energy-efficient {UAV} communication with trajectory
  optimization,'' \emph{IEEE Transactions on Wireless Communications}, vol.~16,
  no.~6, pp. 3747--3760, June 2017.

\bibitem{Baidya/etal/2018a}
S.~Baidya, Z.~Shaikh, and M.~Levorato, ``{FlyNetSim}: {A}n open source
  synchronized {UAV} network simulator based on ns-3 and ardupilot,'' in
  \emph{Proceedings of the 21th ACM International Conference on Modelling,
  Analysis and Simulation of Wireless and Mobile Systems}, ser. MSWiM
  '18.\hskip 1em plus 0.5em minus 0.4em\relax Montreal, Canada: ACM, Oct 2018.

\bibitem{Zema/etal/2018a}
N.~R. Zema, A.~Trotta, E.~Natalizio, M.~Di~Felice, and L.~Bononi, ``The
  {CUSCUS} simulator for distributed networked control systems,'' \emph{Ad Hoc
  Netw.}, vol.~68, no.~C, pp. 33--47, Jan. 2018.

\bibitem{Hadiwardoyo/etal/2019a}
S.~A. {Hadiwardoyo}, C.~T. {Calafate}, J.~{Cano}, Y.~{Ji},
  E.~{Hernández-Orallo}, and P.~{Manzoni}, ``Evaluating {UAV}-to-car
  communications performance: {F}rom testbed to simulation experiments,'' in
  \emph{2019 16th IEEE Annual Consumer Communications Networking Conference
  (CCNC)}, Jan 2019, pp. 1--6.

\bibitem{Sliwa/etal/2017a}
B.~Sliwa, J.~Pillmann, F.~Eckermann, L.~Habel, M.~Schreckenberg, and
  C.~Wietfeld, ``Lightweight joint simulation of vehicular mobility and
  communication with {LIMoSim},'' in \emph{IEEE Vehicular Networking Conference
  (VNC)}, Torino, Italy, Nov 2017.

\bibitem{Sommer/etal/2011a}
C.~{Sommer}, D.~{Eckhoff}, R.~{German}, and F.~{Dressler}, ``A computationally
  inexpensive empirical model of {IEEE 802.11p} radio shadowing in urban
  environments,'' in \emph{2011 Eighth International Conference on Wireless
  On-Demand Network Systems and Services}, Jan 2011, pp. 84--90.

\bibitem{Sliwa/etal/2017b}
B.~Sliwa, J.~Pillmann, F.~Eckermann, and C.~Wietfeld, ``{LIMoSim}: {A}
  lightweight and integrated approach for simulating vehicular mobility with
  {OMNeT++},'' in \emph{OMNeT++ Community Summit 2017}, Bremen, Germany, Sep
  2017, {Best Contribution Award}.

\bibitem{Treiber/etal/2000a}
M.~Treiber, A.~Hennecke, and D.~Helbing, ``Congested traffic states in
  empirical observations and microscopic simulations,'' \emph{Phys. Rev. E},
  vol.~62, no. cond-mat/0002177, pp. 1805--1824, 2000.

\bibitem{Reynolds/1999a}
C.~W. Reynolds, ``{Steering behaviors for autonomous characters},'' \emph{Game
  developers conference}, 1999.

\bibitem{Luukkonen/2011a}
T.~Luukkonen, ``Modelling and control of quadcopter,'' Aalto University, Espoo,
  Tech. Rep., 2011.

\bibitem{Liu/etal/2017a}
Z.~Liu, R.~Sengupta, and A.~Kurzhanskiy, ``A power consumption model for
  multi-rotor small unmanned aircraft systems,'' in \emph{2017 International
  Conference on Unmanned Aircraft Systems (ICUAS)}, June 2017, pp. 310--315.

\bibitem{Falkenberg/etal/2018a}
R.~Falkenberg, B.~Sliwa, N.~Piatkowski, and C.~Wietfeld, ``Machine learning
  based uplink transmission power prediction for {LTE} and upcoming {5G}
  networks using passive downlink indicators,'' in \emph{2018 IEEE 88th
  Vehicular Technology Conference (VTC-Fall)}, Chicago, USA, Aug 2018.

\bibitem{Yacef/etal/2017a}
F.~Yacef, N.~Rizoug, O.~Bouhali, and M.~Hamerlain, ``Optimization of energy
  consumption for quadrotor {UAV},'' in \emph{International Micro Air Vehicle
  Conference and Flight Competition 2017}, H.~d.~P. J.-M.~Moschetta,
  G.~Hattenberger, Ed., Toulouse, France, Sep 2017, pp. 215--222.

\bibitem{Eckermann/etal/2019a}
F.~Eckermann, M.~Kahlert, and C.~Wietfeld, ``Performance analysis of {C-V2X}
  mode 4 communication introducing an open-source {C-V2X} simulator,'' in
  \emph{2019 IEEE 90th Vehicular Technology Conference (VTC-Fall)}, Honolulu,
  Hawaii, USA, Sep 2019.

\bibitem{Seo/etal/2016a}
H.~Seo, K.~Lee, S.~Yasukawa, Y.~Peng, and P.~Sartori, ``{LTE} evolution for
  vehicle-to-everything services,'' \emph{IEEE Communications Magazine},
  vol.~54, no.~6, pp. 22--28, June 2016.

\end{thebibliography}

\end{document}